\documentstyle[12pt,preprint]{aastex}
\newcommand{\beq}{\begin{equation}}
\newcommand{\eeq}{\end{equation}}
\newcommand{\beqn}{\begin{eqnarray}}
\newcommand{\eeqn}{\end{eqnarray}}
\newcommand{\non}{\nonumber}
\newcommand{\pa}{\partial}

\begin{document}

\title{\bf{On the Fundamental
Mass-Period Functions of Extrasolar Planets}}

\author{Ing-Guey Jiang$^a$, Li-Chin Yeh$^b$, Yen-Chang Chang$^b$,
 Wen-Liang Hung$^c$}

\affil{{\small $^a$Department of Physics and Institute of Astronomy,\\
National Tsing Hua University, Hsin-Chu, Taiwan}\\
{\small $^b$Department of Applied Mathematics,\\
National Hsinchu University of Education, Hsin-Chu, Taiwan} \\
{\small $^c$Graduate Institute of Computer Science,\\
National Hsinchu University of Education, Hsin-Chu, Taiwan}}

\begin{abstract}
Employing a catalog of 175 extrasolar planets (exoplanets) detected by the
Doppler-shift method, we constructed the independent and coupled
mass-period functions. It is the first time in this field that the
selection effect is considered in the coupled mass-period
functions. Our results are consistent with those in
Tabachnik and Tremaine (2002) with the major differences that
we obtain a flatter mass function but a steeper period function.
Moreover, our coupled mass-period functions show that about 2.5 percent 
of stars would have a planet with mass between Earth Mass and Neptune 
Mass, and about 3 percent of stars
would have a planet with mass between Neptune Mass and Jupiter
Mass ($M_J$).
\end{abstract}


\section{Introduction}

The mass (size) function, i.e. the differential form for the number of objects
as a function of mass (size),
\beq
f_M(m) = {\frac{dN}{dm}},
\eeq
has been an important physical property
to be investigated in many fields
of astronomy such as galaxies, stars, asteroids and also dust grains.
The importance lies in the meaning that this function is related
with the formation and evolution of particular types of objects.
Due to the fact that the mass function can be studied either
through observational techniques or theoretical calculations,
numerous research projects have been done on this subject.

For stars, the initial mass function (IMF) is the distribution of
stellar masses from one star formation event in a given volume of
space. Although the star-forming conditions vary with the
environment, the measured IMF appears to be universal and can be
modeled by a power-law, \beq f_M(m)={\frac{dN}{dm}}=c_{\star}\
m^{-\alpha_{\star}}, \eeq where $c_{\star}$ is a normalization
constant, and $\alpha_{\star}=2.35$ for the well-known Salpeter
IMF. According to Kroupa, Tout, and Gilmore (1993), \beq\left\{
\begin{array}{ll}
& f_M(m)= c_1 m^{-4.5}\ \ {\rm for}\ \ m > 1 M_{\odot},\\
& f_M(m)= c_2 m^{-2.2}\ \ {\rm for}\ \ 0.5 M_{\odot} < m < 1 M_{\odot},\\
& f_M(m)= c_3 m^{-1.2}\ \ {\rm for}\ \  m < 0.5 M_{\odot},
\end{array}\right.
\eeq
where $m$ is the star's mass, and
$c_1$, $c_2$, $c_3$ are constants to be determined by the total number of
stars in the considered system.

Moreover, there is a new development in astronomy that more than
300 planets are detected around solar type stars. The discovery
has led to a new era of the study of planetary systems and thus
triggered many interesting or controversial results of theoretical
works (Jiang \& Ip 2001, Kinoshita \& Nakai 2001, Armitage et al.
2002, Ji et al. 2003, Boss 2005, Jiang \& Yeh 2007, Rice et al.
2008, Ji et al. 2009). For example, many discovered exoplanets
have extremely short orbital periods. It is likely that they are
formed at larger radial distances and migrate to the current
locations later. However, because the migration timescale is too
short, the rapid inward Type I migration caused by
disc-core interactions  poses a serious issue.

At the time when there were only about 70 detected exoplanets,
Tabachnik \& Tremaine (2002) first used the maximum likelihood
method to determine the mass and period functions with the
assumption of two independent power laws. This work is, in fact,
the only one that takes into account the selection effect and has
intentions to obtain the fundamental mass-period functions.
We have to note that different from the definition of stellar IMF,
the planetary functions are constructed through the data of
exoplanets from all different systems.

Without considering the selection effect, Zucker \& Mazeh (2002)
calculated the linear correlation coefficient between mass and
period and concluded that the mass-period correlation exists. The
study of correlations within sub-groups in the cluster analysis
done by Jiang et al. (2006) and Marchi (2007) also confirms the
mass-period correlation. Thus, strictly speaking, the mass and
period functions shall not be two independent power laws.
Motivated by the above results, Jiang et al. (2007) employed an
algorithm to generate a pair of positively correlated
$\beta$-distributed random variables in order to construct a
coupled mass-period function, and was the first one in the
field of exoplanets to include the correlation into the
construction of mass-period functions.

Moreover, through a nonparametric approach, Jiang et al. (2009)
further constructed new coupled mass-period functions from 279
exoplanets and presented two main statistic results: (1)
confirming the deficit of massive close-in planets; (2)
discovering that more massive planets have larger ranges of
possible semimajor axes. However, due to the fact that the selection effect
is not considered in the above study, it is unclear  how strong
the statement (2) is. Yeh et al. (2009) argued that the planets
larger than 1 $M_J$ should be all within the detection limit, and
thus implied that (2) should be a statistically valid statement.

Therefore, in order to address the above problem,
it would be a great success
if the coupled mass-period function
can be obtained while the selection effect is considered simultaneously.
It is this goal that motivates this work.
The coupled mass-period function will be constructed by the same statistic
method as in Jiang et al. (2009) and the selection effect
would be considered through the similar procedure used
in Tabachnik \& Tremaine (2002).
However, as this work is an extension from
Tabachnik \& Tremaine (2002), power laws would be employed
as standard forms of mass-period functions.
In order to consider the selection effect, only
the exoplanets detected by the Doppler-shift method will be included
as our samples.

We first constructed a reference-based catalog in \S 2, and
discussed the normalization problem in \S 3. Then, the independent
mass-period function was constructed in \S 4.  In \S 5,
mass-period correlations was studied. The coupled mass-period
function was established in \S 6, and the fraction of stars with
planets was discussed in \S 7.  We concluded the paper in \S 8.

\section{A Reference-Based Catalog}

The most well-known catalog of exoplanets is the one maintained by Jean
Schneider (Schneider Catalog hereafter),
i.e. http://exoplanet.eu/catalog.php. This catalog is updated
frequently when new detections are reported in refereed papers or
conference proceedings.
It is also notable that Butler et al.(2006a) updated the orbital solutions
and compiled a list of 172 exoplanets.
In order to extend the work of Tabachnik and Tremaine (2002) with
more updated data, we here constructed a new exoplanet catalog, in
which all samples were discovered by the Doppler-shift method, as
listed in Appendix A. A major principle to construct this catalog
is that all exoplanets in our catalog shall be reported as new
detections in papers of refereed journals.
In order to establish such a reference-based catalog, we searched
and reviewed many published papers. We intended to make sure that
these papers did report new discoveries, and to check which
observational survey the results belong to. Thus, all the
references listed in our catalog are the papers which reported new
detections.


In our catalog, the first column is the data-set identity and the
2nd column is the name of the observational survey. The 3rd and
4th columns are the reference papers and the papers' corresponding
identities.  The 5th column lists the number of exoplanets
discovered in that corresponding paper. Finally, column 6, 7, and
8 give the name, the projected mass $M$, and the orbital period
$P$ of exoplanets. Most planets' $M$ and $P$ are obtained from
Schneider Catalog. The values in Butler et al. (2006a) are used
when they are missing in Schneider Catalog. We have to note that the
exoplanet HD 154345b in Ref. (E-1) is removed  here due to its
extraordinarily larger period, 10900 days.

Besides, we found that some exoplanets were detected by more than
one group around the same time, they could be reported as new
discoveries in two different papers. We checked these and made a
list in Table 2. However, they are still repeatedly listed in
Appendix A, as our catalog was constructed based on published
papers. Moreover, some papers studied more than one exoplanets,
among which only one planet is a new discovery. In those papers,
additional planets were included for a comparison purpose. We also
carefully examined these kinds of papers, so that the reference
papers in the catalog are exactly the papers which discover those
listed exoplanets.

\section{The Normalization Problem}

In addition to the selection effect, the number of observed stars
in a survey, $N_s$, is one of the key parameters to understand the
probability that a star could host an exoplanet with particular
orbital properties. Unfortunately, we found that the exact numbers
of stars in surveys were not clearly mentioned normally. In most
cases, an approximate number of target stars in a continuous
long-term survey might be given, but the exact number of observed
stars at that time when a new exoplanet detection was detected was
usually not stated. For example, in Butler et al. (2003), it is
written as ``The Keck survey includes about 650 main-sequence and
subgiant stars ...'', and later in the same section, it is stated
as ``200 stars have one or more Keck observations but have been
subsequently dropped from the program ...''. In this case, neither
650 nor 450 can be used as the $N_s$ here.

Fortunately, Lineweaver and Grether (2003) obtained the numbers of
target stars in surveys carefully and listed the results in their
Table 4. In fact, they also estimated the number of repeated stars
in different surveys, and finally determined the total number of
target stars to be 1812 in these surveys. The numbers of detected
planets in the corresponding surveys are also shown in their Table
3, and the total number is 122. Therefore,  we decide to set
$N_{\rm ratio}\equiv N/N_{s} \equiv 122/1812=0.0673$ in our paper.
This value of $N/N_{s}$ would be used when we determine the value
of the normalization constant $c$ through the related equations,
and also as a way to obtain $N_{s}$ in a particular survey for a
given number of detected exoplanet $N$.

\section{Independent Mass-Period Functions}

In this section, the method and results based on the assumption of
independent mass and period power-law functions will be described.

\subsection{The Method}
The procedures in this section, i.e. the analytical approach, the
choice of parameters etc., follow exactly that of Tabachnik and
Tremaine (2002). Here we describe the method in a self-consistent
way. However, please note that we simply consider the projected
(minimum) mass $M$, which satisfies $M=M_{\rm real} \sin i$, where
$M_{\rm real}$ is the real physical mass of the exoplanet and $i$ is
the orbital inclination angle, in all calculations in this paper.
That is, the mass $M$ in this paper means the minimum mass. The
probability, $dp$, that a single star has an exoplanet with mass $M$
and orbital period $P$ in the range, $[M, M+dM]$, $[P, P+dP]$, is
given by the product of independent power laws on $M$ and $P$ as:
\beq dp=c \left[\frac{M}{M_0}\right]^{-\alpha}
\left[\frac{P}{P_0}\right]^{-\beta} \frac{dM}{M}\frac{dP}{P}, \eeq
where $c$, $\alpha$ and $\beta$ are constants to be determined, and
$M_0=1.5 M_{J}$, $P_0= 90$ days. We assume that there are $N$
exoplanets in the data set, and let \beq\left\{
\begin{array}{ll}
& x_{i}=\ln \left(\frac{M_i}{M_0}\right)\\
& y_{i}=\ln \left(\frac{P_i}{P_0}\right)\quad {\rm for} \quad 1\le i\le N \\
\end{array}\right.
\eeq
where $M_i$, $P_i$ are the mass
and orbital period of one particular exoplanet.
According to Eq.(2) and (6) of Tabachnik and Tremaine (2002),
the value of $x_i - y_i/3$ shall satisfy
\beq
x_i - y_i/3 \ge {\rm ln}\left[ \frac{K_D}{28.4 {\rm m s^{-1}} } \right]
             -{\rm ln}\left[ \frac{M_0}{M_J} \right]
             + \frac{1}{3}{\rm ln} \left[ \frac{P_0}{1{\rm yr}} \right],
\eeq
where $K_D$ is the detection limit of the considered survey.
For convenience, we define the right hand side of the above equation to be
\beq
v_D \equiv  {\rm ln}\left[ \frac{K_D}{28.4 {\rm m s^{-1}} } \right]
             -{\rm ln}\left[ \frac{M_0}{M_J} \right]
             + \frac{1}{3}{\rm ln} \left[ \frac{P_0}{1{\rm yr}} \right].
\eeq The value of $v_D$ can be determined if the detection limit
$K_D$ is exactly known. However, a range of possible values of
$K_D$ is usually stated in a paper because it is related with the
condition of instruments and weather. Similar to Tabachnik and
Tremaine (2002), we set the smallest value of $x_i - y_i/3$, i.e. 
$ \min_{1\le i\le N} \left\{x_i- y_i/3\right\} $,
 in the
considered data to be the value of $v_D$ and obtain the most likely
$K_D$ reflected from the data through the above equation. Further,
any value of $y_i$ shall satisfy 
\beq y_i \le {\rm ln} \left[
\frac{P_{\rm max}}{P_0} \right], 
\eeq 
where $P_{\rm max}$ is the
upper limit of orbital periods of detectable systems, which is
proportional to the duration of the survey, as at least two orbits
are required for a reliable detection. For convenience, we define
\beq 
u_D \equiv {\rm ln} \left[ \frac{P_{\rm max}}{P_0} \right].
\eeq 
Similarly, it is not easy to obtain a consistent $P_{\rm
max}$ from the published papers. We set the largest value of $y_i$
in the considered samples to be the value of $u_D$ and get the most
likely $P_{\rm max}$ reflected from the data.

On the other hand, there are further constraints on $x_i$ and $y_i$.
Let $M_{\rm max}$ be the maximum value of the mass
in the data set for a considered survey, and define
\beq
x_{\rm max} = {\rm ln}  \left[ \frac{M_{\rm max}}{M_0} \right],
\eeq
so we have a constraint on $x_i$ as
\beq
x_i \le x_{\rm max}.
\eeq
Similarly, let $P_{\rm min}$ be the minimum value of the orbital period
in the data set for a considered survey, and define
\beq
y_{\rm min} = {\rm ln}  \left[ \frac{P_{\rm min}}{P_0} \right],
\eeq
we have another constraint on $y_i$ as
\beq
y_i \ge y_{\rm min}.
\eeq

Therefore, in the $x-y$ space, the exoplanet probability $dp$
can be expressed as:
\beq
dp = c \left[{\rm e}^x \right]^{-\alpha}
       \left[{\rm e}^y \right]^{-\beta} dx dy,
\eeq and the expected number of exoplanets in the area $dx dy$ in a
survey of $N_{s}$ stars can be written as \beq n(x,y) dx dy = N_{s}
dp = c N_{s} {\rm e}^{-\alpha x}
       {\rm e}^{-\beta y} dx dy.
\eeq
Thus, as in Tabachnik and Tremaine (2002),
the likelihood function $L$ is expressed as
 \beq
 L=\Pi^{N}_{i=1} n(x_i,y_i)\exp [-\int_D n(x,y) dxdy],\label{eq:l}
 \eeq
where the domain $D$ is 
$v+y/3 \le x \le x_{\rm max}, y_{\rm min} \le y \le \tilde{u}$ with
\beq\left\{
\begin{array}{ll}
& v=\min_{1\le i\le N} \left\{x_i-\frac{y_i}{3}\right\}\\
& u=\max_{1\le i\le N}\left\{y_i\right\} \\
&\tilde{u}=\min\left\{u,3(x_{\max}-v)\right\}
\end{array}\right.
\eeq
Finally, through the maximum likelihood method, the values of $c$,
$\alpha$, and $\beta$ can be determined by the following
equations:
\beq \frac{\pa \ln L}{\pa c}=0, \quad \frac{\pa \ln
L}{\pa \alpha}=0, \quad {\rm and}\  \frac{\pa \ln L}{\pa
\beta}=0.\label{eq:ln_l}
\eeq
The first one, ${\pa \ln L}/{\pa c}=0$, can be solved analytically
and yielded an analytical expression of $c$ as
shown in Eq. (16) of Tabachnik and Tremaine (2002).
Using this analytical expression of $c$, ${\pa \ln L}/{\pa \alpha}=0$
and ${\pa \ln L}/{\pa \beta}=0$ can be solved numerically.

\subsection{Individual Surveys}

First, the mass-period function is estimated through the data from
one given particular survey.
The previously mentioned equations are used to obtain the
estimates of $\alpha$, $\beta$, and $c$. Their error bars are
calculated as the standard deviations via the bootstrap algorithm
with the replication size $B=2000$ (Jiang et al. 2007). The
results are shown in Table 1, and each row is for one particular
survey.

\vskip 0.2truein
\smallskip

 \centerline{ {\bf Table 1.} }
 \begin{center}
  \begin{tabular}{|l|l|c|c|c|c|c|c|}\hline
  Survey  & Data Set  & N & $\alpha$  & $\beta$
& $c\times 10^{3}$ & ${\rm K_D(ms^{-1})}$ &${\rm P_{max}}$(yr)
       \\ \hline
Lick&(A)& 7& -0.401$\pm$1.051 & -0.144$\pm$0.388  & 4.466 $\pm$  4.915  &
  19.794  & 14.296  \\\hline
Coralie&(B)& 38 &  0.332$\pm$0.186 & -0.393$\pm$0.101 &  2.969 $\pm$  0.372
& 21.032  & 5.759\\\hline
Elodie& (C)&14& -0.473$\pm$0.827 & -0.343$\pm$0.291& 2.327 $\pm$  5.876
 & 21.400  & 7.921\\\hline
HARPS &(D) &23&  0.081$\pm$0.146  & 0.097$\pm$0.146 & 1.733 $\pm$  4.189  &
1.128  & 2.844\\\hline
N2K &(E) &14& -0.071$\pm$0.425 & 0.077$\pm$0.196 & 3.320 $\pm$ 4.057 &
  12.998& 3.507\\\hline
Keck &(F)& 48& 0.131$\pm$0.135 & -0.326$\pm$0.095 & 1.362 $\pm$ 0.243 &
  2.9304& 9.184 \\\hline
AAPS&(G) &23& 0.296$\pm$0.262& -0.558$\pm$0.202 & 1.778 $\pm$  0.715 &
   10.318 &8.181 \\\hline
Other&(H) &17& 0.168$\pm$0.350  &-0.341$\pm$0.192 & 3.896 $\pm$  3.154 &
  28.189 & 5.957\\ \hline
Single& & 175& -0.143$\pm$0.040& -0.124$\pm$0.044&  1.316 $\pm$  0.080 &
  1.128& 14.296 \\ \hline
         \end{tabular}
   \end{center}

The data sets and total numbers of exoplanets used in the
calculation of maximum likelihood method are shown in columns 2 and
3. The resulting values of $\alpha$ and $\beta$, i.e. the exponents
of the power-law mass-period functions, are in columns 4 and 5. To
make it clear, their values with error bars are plotted in Fig. 1.
Table 1 shows that the Lick results have the largest error bars, due
to the small numbers of planets. Note that Lick Survey has the
smallest number of planets here. The reason is that we put the
planets in Fischer et al.(2001) into Other Survey because both data
came from Keck and Lick telescopes are used in that paper. We found
that, among these surveys, the Coralie and Keck results have smaller
error bars as Coralie Survey has 38 and Keck Survey has 48 planets.
HARPS Survey has 23 planets, but the error bars are about as small
as those in  Coralie Survey due to the high precision on the
measurement of radial velocities. In general, the absolute values of
$\alpha$ are less than 0.5, and the values of $\beta$ range from
-0.6 to 0.1. They both can be positive or negative.

The final row is the result when all the above exoplanets are assumed
to be discovered by a Single Imaginary Survey. However, as shown in Table 2,
some planets are included in more than one data set.
Although the summation of numbers of planets from all data sets is 184,
there are actually 175 planets in total.
The $\alpha$ and $\beta$ of the result of
Single Imaginary Survey
are both negative, and the corresponding
error bars are very small due to a much larger number
of samples.\\

\centerline{ {\bf Table 2. Repeated Exoplanets} }
 \begin{center}
  \begin{tabular}{|c|l|l|}\hline
 &  Planet       & Reference ID \\ \hline
(1) & HD102117 & (G-12) and (D-5)\\ \hline
(2) & HD 196050& (B-1) and (G-2)\\ \hline
(3)& HD 216437& (B-1) and (G-2)\\ \hline
(4) & HD 52265 &(B-4) and (F-16) \\ \hline
(5)& HD 192263 & (B-7) and (F-7)\\ \hline
(6)&  HD 168443 c&(B-11) and (F-10)\\ \hline
(7)& HD 33636 &(C-2) and (F-9) \\ \hline
(8)&  HD 37124 c&(F-1) and (F-12)\\ \hline
(9)&  HD 92788 & (B-1) and (H-1)\\ \hline
       \end{tabular}
    \end{center}

\subsection{Multiple Surveys}

The analysis of individual surveys is generalized to
Multiple Surveys here.
In this analysis, different domains are considered for
different surveys in Eq. (\ref{eq:l}).  Thus, from
Eq.(\ref{eq:l}), the likelihood function $L_j$ for the survey $j$ is
 \beq
 L_j=\Pi^{N_j}_{i=1} n_j(x_{i,j},y_{i,j})\exp [-\int_{D_j} n_j(x,y) dxdy],
\label{eq:lj}
 \eeq
where $N_j$ is the number of discovered planets for the survey $j$, and
$$n_j(x,y)=c N_{s}e^{-{\alpha} x -{\beta} y}
=\frac{c N_j}{N_{\rm ratio}}
e^{-{\alpha} x -{\beta} y}.
$$
Thus, the log-likelihood is given by \beq \ln
{L}=\sum^{J}_{j=1}\ln L_j, \eeq where $J$ is total number of
considered surveys, and we have $J=8$ here. When $\ln L$ is
maximized, the estimates for the parameters are obtained as: \beqn
& &{\alpha}= -0.099 \pm  0.055\\
& &{\beta}= -0.129  \pm 0.040\\
& &{c} = ( 2.423 \pm 0.154 )\times 10^{-3}, \eeqn where the error
bars are also estimated via the bootstrap algorithm (Jiang et al.
2007). This set of $\alpha$ and $\beta$ with error bars are also
shown in Fig. 1.

In order to show the distributions implied by these results and to
be compared with the samples' histograms, we take integrations for
the function $n(x,y)$, and thus the number of planets  with masses
between $M_1$ and $M_2$ and periods between $P_1$ and $P_2$ is
given by
\beqn
N_{[M_1,M_2][P1,P2]}&=&
\int^{\ln (P_2/P_0)}_{\ln (P_1/P_0)} \int^{\ln (M_2/M_0)}_{\ln (M_1/M_0)}
n(x,y)dxdy=\int\int c N_{s} e^{-\alpha x-\beta y}dxdy, \non \\
&=& -\frac{c N_{s}}{\alpha}\int^{\ln (P_2/P_0)}_{\ln (P_1/P_0)}
\left\{\left(\frac{M_2}{M_0}\right)^{-\alpha}-\left(\frac{M_1}{M_0}\right)
^{-\alpha}\right\} e^{-\beta y}dy \non  \\
&=& \frac{c
N_{s}}{\alpha\beta}\left\{\left(\frac{M_2}{M_0}\right)^{-\alpha}
-\left(\frac{M_1}{M_0}\right)^{-\alpha}\right\}\left\{\left(\frac{P_2}{P_0}
\right)^{-\beta}-\left(\frac{P_1}{P_0}\right)^{-\beta}\right\}
\eeqn After substituting the result of Single Imaginary Survey
into the above equation, we obtain the expected number of planets
in given mass or period intervals as shown in Fig. 2. With mass
interval 0.5 Jupiter-mass and period interval 150 days, the
crosses connected with dotted lines in Fig. 2 show the planetary
distributions in mass space (see Fig. 2(a)),
and also in period space (see Fig. 2(b)).
For the comparison, the mass and period histograms of 175 exoplanets
and the corresponding results derived from the
Multiple Surveys are also plotted
in Figs. 2(a)-(b).

\section{Mass-Period Correlation Coefficients}

In addition to the mass and period histograms and
distributions shown in Fig. 2,
the exoplanets' locations on the mass-period space, i.e. $M-P$ space,
are often presented,
as the mass
and orbital period are the most important physical parameters
of exoplanets.
Fig. 3 is the distribution of 175 exoplanets
in logarithmic space, $x-y$ space,
where $x, y$ are as defined previously.
The region enclosed by four solid lines is the Domain D,
where all 175 planets are included.
In order to investigate the strength of mass-period correlations,
the Spearman rank-order correlation coefficients $\rho_{S}$
are calculated,
as shown in Table 3.\\

 \centerline{ {\bf Table 3.} }
 \begin{center}
  \begin{tabular}{|l|c|c|c|}\hline
  Survey& Data Set & $N$&   $\rho_{S}$\\ \hline
 Lick&(A)&7&   0.143  \\ \hline
Coralie&(B) & 38&  0.351   \\ \hline
Elodie &(C)&14&  0.431    \\ \hline
HARPS &(D)  &23&  0.564  \\ \hline
N2K &(E)&14&  0.552  \\ \hline
Keck &(F) & 48&  0.532  \\ \hline
AAPS &(G) &23&   0.553 \\ \hline
Other&(H) &17&   0.542  \\ \hline
Single & & 175&  0.540   \\ \hline
         \end{tabular}
\end{center}

In general, the values
of $\rho_{S}$ are larger than 0.5 in five individual surveys and
the Single Imaginary Survey of
overall 175 planets. These results imply a strong positive
correlation (Cohen 1988).
Because there is a mass-period correlation, we
shall consider the mass-period coupling and
construct coupled mass-period functions in this paper.

On the other hand, as pointed in Jiang et al.(2006) and in
Marchi (2007), the correlation of each group of exoplanets identified
in the clustering analysis might be linked with the physical mechanisms
more easily. The overall correlations could be more difficult to explain.
A recent result by Marchi et al. (2009) is a good example that two dominated
groups of close-in exoplanets are found and can be explained by
two physical mechanisms successfully. This is an interesting and important
topic that we would like to do further investigations in the future.
However, the main point in this section is to re-confirm that there is indeed
a mass-period correlation for our selected samples, and thus it is necessary
to construct the coupled mass-period functions in this paper.

\section{Coupled Mass-Period Functions}

We consider the mass-period coupling and
construct coupled mass-period functions here.
We define variables $x=\ln(M/M_0)$, $y=\ln(P/P_0)$,
and their probability density functions as:
\beqn
& &f_X(x)=\frac{\alpha e^{-\alpha x}}{e^{-\alpha x_{\rm min}}-e^{-\alpha x_{\rm max}}},~
     x_{\rm min} \le x \le x_{\rm max};\\
& &f_Y(y)=\frac{\beta e^{-\beta y}}
{e^{-\beta y_{\rm min}}-e^{-\beta y_{\rm max}}},~
     y_{\rm min} \le y \le y_{\rm max}.
 \eeqn
Based on the copula modeling method introduced in Jiang et al. (2009),
the coupled probability density function
$f_{XY}(x,y)$ is
\begin{eqnarray}
& & f_{XY}(x,y) =\frac{\partial^2 C (F_X(x),F_Y(y);\theta)}
      {\partial F_X \partial F_Y} f_X(x) f_Y(y), \non \\
&=&\frac{-\theta(e^{-\theta}-1)
 e^{-\theta F_X(x)}e^{-\theta F_Y(y)}}
 {\Big[e^{-\theta}-1+(e^{-\theta F_X(x)}-1)(e^{-\theta F_Y(y)}-1)\Big]^2}
f_X(x)f_Y(y),\label{eq:f_pm}
\end{eqnarray}
where the function $C (F_X(x),F_Y(y);\theta)$ is given by
\beq
 C(u_1,u_2;\theta)=\frac{-1}{\theta} \ln\Big[
  1+\frac{(e^{-\theta u_1}-1)(e^{-\theta u_2}-1)}{e^{-\theta}-1}\Big],
\eeq
$u_1$ is the integral of $f_X(x)$, and $u_2$ is
the integral of $f_Y(y)$. Thus,
\begin{eqnarray}
 u_1 &\equiv& F_X(x) \equiv \frac{e^{-\alpha x_{\rm min}}-e^{-\alpha x}}
 {e^{-\alpha x_{\rm min}}-e^{-\alpha x_{\rm max}}},~
x_{\rm min} \le x \le x_{\rm max},\\
 u_2 &\equiv& F_Y(y) \equiv \frac{e^{-\beta y_{\rm min}}-e^{-\beta y}}
 {e^{-\beta y_{\rm min}}-e^{-\beta y_{\rm max}}},~
y_{\rm min} \le y \le y_{\rm max}.
\end{eqnarray}
The dependence parameter
$\theta$ ($-\infty<\theta<\infty$)
can be positive, zero and negative,
corresponding to the positive dependence, independence and
negative dependence between two variables $x$ and $y$, respectively.
When $\theta$ approaches to zero, the term
$\partial^2 C/\partial F_X \partial F_Y$ would approach to one,
and $f_{XY}(x,y)= f_X(x)f_Y(y)$. Thus,
$\partial^2 C/\partial F_X \partial F_Y$
is called the {\it coupling factor} in this paper as it controls
the $x-y$ dependence.

The function $C(u_1,u_2;\theta)$ in 
Eq. (28) is called the Frank copula function.
In fact, there are many available copula functions, and the reason why
we choose this one is that it is more flexible as it allows to have
negative, zero, and positive correlations.


Further, the expected number of exoplanets in the area $dx dy$
in a survey of $N_s$ stars is
\beq
n(x,y) dx dy = c N_{s} f_{XY}(x,y) dx dy, \label {eq:n_cp}
\eeq
where the parameter $c$ is a constant to be determined.
For a given function $n(x,y)$,
the number of planets  with masses between
$M_1$ and $M_2$ and periods between $P_1$ and $P_2$ is determined by
\beqn
N_{[M_1,M_2][P1,P2]}
&=&
\int^{\ln (P_2/P_0)}_{\ln (P_1/P_0)} \int^{\ln (M_2/M_0)}_{\ln (M_1/M_0)}
n(x,y)dxdy \non \\
&=& c N_{s}
\int^{\ln (P_2/P_0)}_{\ln (P_1/P_0)} \int^{\ln (M_2/M_0)}_{\ln (M_1/M_0)}
f_{XY}(x,y)dxdy.
\eeqn
On the other hand, the likelihood function is
\beqn
L &=& \prod_{i=1}^N  n(x_i,y_i)
\exp\Big[-\int_D n(x,y)dxdy\Big], \non \\
&=&\prod_{i=1}^N c  N_{s} f_{XY}(x_i,y_i)
\exp\left[-cN_s \int_{y_{\rm min}}
^{\tilde{u}}
\int_{v+y/3}^{x_{\rm max}} f_{XY}(x,y)dxdy\right].
\eeqn

After the derivation shown in Appendix B, $\ln L$ is finally expressed
as a function of $\alpha$, $\beta$, and $\theta$.
The maximum likelihood method is used to simultaneously estimate
the parameters $\alpha, \beta$ and $\theta$ through the full log
likelihood $\ln L$.
The estimates of $c$, $\alpha$, $\beta$ and $\theta$ for each survey
are listed in Table 4. Moreover, as the procedure in \S 4, here we
also generalize the result to the case of Multiple Surveys and the
result is at the bottom of Table 4. The bootstrap method is also
used to get error bars. Fig. 4 shows the values of $\alpha$ and
$\beta$ with error bars. The result of Single Imaginary Survey of
175 planets gives $\alpha= -0.187\pm 0.034$, $\beta=-0.133\pm
0.041$, so the mass (period) function has a power-index $-0.813\pm
0.034$ ($-0.867\pm 0.041$). Moreover, the result of Multiple Surveys
gives $\alpha = -0.038\pm 0.080$, $\beta = -0.137\pm 0.044$, and
thus the mass (period) power-index is $-0.962\pm 0.080$ ($-0.863\pm
0.044$). On the other hand, the mass (period) power-index obtained
in Tabachnik and Tremaine (2002) is $-1.11\pm 0.10$ ($-0.73\pm
0.06$). Thus, considering the error bars, our results are consistent
with those in Tabachnik and Tremaine (2002). However, our mean
values imply a flatter mass function but a slightly steeper period
function. We hope to use the future data to investigate whether this
new result remains to be valid.

In Fig. 5(a), from the result of Multiple Surveys, the coupled
probability density function in $x-y$ space, $f_{XY}(x,y)$, is shown
as a three-dimensional plot. The corresponding contour is in Fig.
5(b). To visualize it in a realistic space, the above two are
transformed to be in $M-P$ space and shown in Figs. 6(a)-(b), where
the coupled probability density function in $M-P$ space,
$f_{MP}(M,P)$ is defined by
 \beq f_{MP}(M,P) = f_{XY}(x,y) |J|, \eeq
where $J$ is the Jacobian determinant 
$\partial x/\partial M  \times \partial y/\partial P$.
As we know from Eq.(27), the mass-period coupling is
primarily determined by the coupling factor. In order to visualize
it, the three-dimensional and color contour plots in $x-y$ space are
shown in Figs. 7(a)-(b). It shows that the coupling factor is bigger
when $x$ and $y$ have the same sign, i.e. both positive or both
negative. This implies that each group of exoplanets would have its
own strength of mass-period correlations if the exoplanets are
clustered into a few groups. For example, Marchi et al. (2009) did
the clustering analysis on exoplanets, and found the existence of
two types of close-in planets. They discovered that these two types
of planets have very different distributions of semi-major axes.
Moreover, Marchi et al. (2009) also proposed possible mechanisms to
produce these two different types of exoplanets.

Figs. 8(a)-(b) show the coupled mass-period probability density
function (pdf), $f_{XY}(x,y)$, for a few given masses or periods.
For the purpose of comparison, the pdfs without coupling factor,
i.e.
 $f_X(x)\times f_Y(y)$
are also plotted in Figs. 8(c)-(d). In Fig. 8(a) and Fig. 8(c),
the solid curve is for $x$=ln($1 M_J/M_0$), the dotted curve is
for $x$=ln($5 M_J/M_0$), the short dashed curve is for $x$=ln($10
M_J/M_0$), and the long dashed curve is for $x$=ln($15 M_J/M_0$).
In Fig. 8(b) and Fig. 8(d), the solid curve is for $y$=ln($1
{\rm day}/P_0$), the dotted curve is for $y$=ln($50 {\rm
days}/P_0$), the short dashed curve is for $y$=ln($100 {\rm
days}/P_0$), and the long dashed curve is for $y$=ln($150 {\rm
days}/P_0$). In general, it shows that the coupling factor does
change the shapes of mass and period functions.
To be more realistic, the curves in Fig. 8 are re-plotted as
functions of $M$ and $P$ as shown in Fig. 9.\\

\clearpage

\centerline{ {\bf Table 4.} }
\begin{center}
  \begin{tabular}{|l|c|c|c|c|c|c|c|}\hline
  Survey& Data Set & N   & $\alpha$ & $\beta$ & $\theta$ &$c\times 10^{2}$
\\ \hline
 Lick&(A)&7&   0.531 $\pm$ 1.047&  -0.124 $\pm$0.334& 2.289 $\pm$ 5.201&
7.520 $\pm$ 6.044 \\ \hline
Coralie&(B) & 38 &  0.314 $\pm$ 0.158&  -0.344$\pm$ 0.096 &
 0.963$\pm$  1.438 & 10.101 $\pm$  1.681 \\ \hline
Elodie &(C)&14&  -0.455 $\pm$ 0.565& -0.203$\pm$ 0.233 &
4.757 $\pm$ 2.978& 6.878 $\pm$  11.234  \\ \hline
HARPS &(D)  &23& 0.046$\pm$ 0.193 &  0.092$\pm$ 0.168 &  1.831 $\pm$ 1.863
 &  7.013 $\pm$  0.312\\ \hline
N2K &(E)&14& 0.045 $\pm$  0.438& 0.130 $\pm$ 0.191 & 1.795$\pm$ 2.045 &
7.049 $\pm$  1.621\\ \hline
Keck &(F) & 48& 0.019$\pm$0.132  & -0.217$\pm$ 0.082 &3.957 $\pm$  1.521 &
 8.145 $\pm$  0.882 \\ \hline
AAPS &(G) &23&  0.018 $\pm$ 0.220 &-0.416$\pm$0.139 &  6.786$\pm$ 2.288&
  7.337 $\pm$  1.353\\ \hline
Other&(H) &17&  -0.069 $\pm$  0.402 & -0.229$\pm$0.182 &  3.970$\pm$ 2.511&
7.166 $\pm$  1.635\\ \hline
Single & & 175&  -0.187 $\pm$ 0.034 & -0.133$\pm$  0.041 &  4.848$\pm$ 0.779
&  6.832 $\pm$  0.098 \\ \hline
Multiple& & 184&  -0.038 $\pm$ 0.080 & -0.137$\pm$  0.044 &  2.973$\pm$ 0.639
&  7.378 $\pm$  0.289 \\ \hline
         \end{tabular}
\end{center}

\section{The Fraction of Stars with Planets}

From both theoretical and observational points of view,
it is important to know what fraction of stars would have planets.
If we divide $N_{[M_1,M_2][P1,P2]}$ by $N_s$ in Eq. (24), the probability
that a star could host a planet with masses between
$M_1$ and $M_2$ and periods between $P_1$ and $P_2$ is given by
\beq
Prob
=\frac{c}{\alpha\beta}\left\{\left(\frac{M_2}{M_0}\right)^{-\alpha}
-\left(\frac{M_1}{M_0}\right)^{-\alpha}\right\}\left\{\left(\frac{P_2}{P_0}
\right)^{-\beta}-\left(\frac{P_1}{P_0}\right)^{-\beta}\right\}
\eeq
for the independent mass-period function.

As the power indexes obtained from the results of Single Imaginary
Survey are different from those from Multiple Surveys, we
calculated both results using Eq. (35). We used the symbol
$Prob_S$ to represent the results of  Single Imaginary Survey and
used $Prob_M$ for the results of  Multiple Surveys.

On the other hand, for coupled mass-period functions, we have
\beqn
& &Prob_{C}={\frac{1}{N_s}}
\int^{\ln (P_2/P_0)}_{\ln (P_1/P_0)} \int^{\ln (M_2/M_0)}_{\ln (M_1/M_0)}
n(x,y)dxdy=\int\int c f_{XY}(x,y)dxdy, \non \\
& =& c\int\int
\frac{-\theta(e^{-\theta}-1)
 e^{-\theta F_X(x)}e^{-\theta F_Y(y)}}
 {\Big[e^{-\theta}-1+(e^{-\theta F_X(x)}-1)(e^{-\theta F_Y(y)}-1)\Big]^2}
f_X(x)f_Y(y) dxdy. \label{eq:nu_cp} \eeqn Similarly, we used the
symbol $Prob_{CS}$ to represent the results of Single Imaginary
Survey and used $Prob_{CM}$ for the results of Multiple Surveys in
the case of coupled mass-period functions.

Tabachnik \& Tremaine (2002) estimated the expected number of
planets per star for given period and mass range through their
best results of multiple surveys, which is in fact equivalently
defined as $Prob_M$ here. When $M_1 = M_J$, $M_2 = 10 M_J$, $P_1=
2$ days, and $P_2= 10$ yrs = 3650 days, Tabachnik \& Tremaine
(2002) found $Prob_M=0.036$ and concluded that 4 percent of
solar-type stars have a planet in the above ranges. For the same
given ranges, we obtain: $Prob_S=0.02618$, $Prob_M=0.04667$,
$Prob_{CS}=0.02909$, and $Prob_{CM}=0.02273$. Thus, our results of
Multiple Surveys are similar to those in Tabachnik \& Tremaine
(2002). Moreover, the estimated probabilities from the coupled
mass-period functions are smaller but still consistent with those
in Tabachnik \& Tremaine (2002). In Tabachnik \& Tremaine (2002),
the case with  $M_1 = 0.003 M_J$ (i.e. Earth Mass), $M_2 = 10
M_J$, $P_1= 2$ days, and $P_2= 10$ yrs = 3650 days is also
estimated and has a probability as 0.18.  However, our results
show that: $Prob_S=0.06406$. $Prob_M=0.1264$, $Prob_{CS}=0.06492$,
and $Prob_{CM}=0.0736$.
Note that the result with $M_1 = 0.003 M_J$ (i.e. Earth Mass)
is an extrapolation as this small mass is out of the mass range of
175 samples. Under the assumption that the smaller planets shall
follow the trend of our mass-period function,
and the fact that the Earth-Mass planet does exist in our
universe, i.e. Solar System, this extrapolation shall lead to a
good estimation.

On the other hand, Naef et al. (2005) also gave estimations about
the fractions of star with planets more massive than 0.5 $M_J$
within three periods: 0.7 percent for period $< 5$ days, 4 percent
for period $< 1500 $ days, and 7.3 percent for period $< 3900 $
days. Correspondingly, using our samples and equations here, for
$M_1=0.5 M_J$, $M_2=M_{\rm max}$, $P_1=P_{\rm min}$, and $P_2= 5$
days (Note that, in 175 samples, the smallest mass $M_{\rm
min}=0.0158$, the largest mass $M_{\rm max}=18.39$, the smallest
period $P_{\rm min}=1.328$, the largest period $P_{\rm max}=5218$,
and the units are $M_J$ and days), we obtained the fraction of
stars with planets to be less than 1 percent.
For $M_1=0.5 M_J$, $M_2=M_{\rm max}$, $P_1=P_{\rm min}$, and $P_2=
1500$ days, we obtained the fraction of stars with planets to be
about 3 to 6 percent.
Moreover, for  $M_1=0.5 M_J$, $M_2=M_{\rm max}$, $P_1=P_{\rm min}$,
and $P_2= 3900$ days,
the results are about 4 to 8 percent.

Finally, we are interested in the possibility to have a planet
with mass between Earth Mass and Neptune Mass, for any period,
so we set
$M_1=0.003$, $M_2= 0.05$, $P_1=P_{min}$, and $P_2=P_{max}$, and get:
$Prob_S$ = 1.604 \%,
$Prob_M$ = 3.649 \%,
$Prob_{CS}$ = 1.263 \%, and
$Prob_{CM}$ = 2.542 \%.
We are also interested in the possibility between
Neptune Mass and Jupiter Mass, for any possible period, thus
we set $M_1=0.05$, $M_2= 1$, $P_1=P_{min}$, and $P_2=P_{max}$, and obtain:
$Prob_S$ = 2.589 \%,
$Prob_M$ = 5.182 \%,
$Prob_{CS}$ = 2.319 \%, and
$Prob_{CM}$ = 3.021 \%.
All the above mentioned results calculated by the equations in this paper
are listed in Table 5.\\

\centerline{ {\bf Table 5.} }
 \begin{center}
  \begin{tabular}{|c|c|c|c|c|c|c|c|c|}\hline
Case & $M_1$ & $M_2$ & $P_1$ & $P_2$ & $Prob_S$ & $Prob_M$ & $Prob_{CS}$ &
$Prob_{CM}$ \\ \hline
1 &  1. &  10.&  2.&  3650. & 0.02618 & 0.04667 & 0.02909 &  0.02273 \\ \hline
2 &  0.003 &  10.&  2.&  3650.& 0.06406 & 0.12640 & 0.06492 & 0.0736 \\ \hline
3 & 0.5  & 18.39 & 1.328 & 5. & 0.00453 & 0.00790 & 0.00154& 0.00169\\ \hline
4 & 0.5  & 18.39 & 1.328 & 1500. & 0.03527 & 0.06255 & 0.03095&0.02661\\ \hline
5 & 0.5  & 18.39 & 1.328 & 3900. & 0.04290 & 0.07631 & 0.04196&0.03658\\ \hline
6 & 0.003 & 0.05 & 1.328 & 5218.0 & 0.01604 & 0.03649 & 0.01263&  0.02542\\
\hline
7 & 0.05 & 1. & 1.328 & 5218.0 & 0.02589 & 0.05182 & 0.02319 & 0.03021\\ \hline
\end{tabular}
\end{center}

\section{Concluding Remarks}

In this paper, several steps have been taken to establish the
fundamental mass-period functions of exoplanets. First of all, by
reading a great number of published papers, we constructed a
reference-based catalog of 175 exoplanets, which were all
discovered through the Doppler-shift method. Employing this
catalog, we determine the independent mass-period functions for
individual surveys and also for the case of Multiple Surveys.
Moreover, the coupled mass-period functions are also constructed
for both individual and Multiple Surveys. The selection effects of
surveys are considered in all the results, and thus it is the
first time in this field that the selection effect is considered
in the coupled mass-period functions.
Our results are consistent with those in
Tabachnik and Tremaine (2002) with the main differences that
our results imply a flatter mass
functions but a steeper period function.

On the other hand, our coupled mass-period functions are used to
predict the possible fractions of stars in given mass and period
ranges. Our results are consistent with previous works, and show
that about 2.5 percent of stars would have a planet with mass
between Earth Mass and Neptune Mass, and about 3 percent of stars
would have a planet with mass between Neptune Mass and Jupiter
Mass.

\section*{Acknowledgment}
We thank the anonymous referee for useful suggestions that
improved the paper enormously.  
We also owe a debt of thanks to Scott Tremaine 
for the communications about his paper, Tabachnik and Tremaine (2002),
and the positive remarks on our paper.
We are grateful to the National Center for High-performance Computing
for computer time and facilities.
This work is supported in part 
by the National Science Council, Taiwan.


\section*{Appendix A}
\begin{center}
\begin{tabular}{|c|l|l|l|c|l|c|c|}\hline
 {Data Set} & Survey & Reference & Ref. ID  & N & Planet &  $M$ & $P$ \\
\hline
 (A) & Lick & Marcy et al.(2002) & (A-1) & 1 & 55 Cnc d & 3.835 & 5218\\ \cline{3-8}
   & & Fischer et al.(2002a) & (A-2) & 1 & 47 Uma c &   1.34& 2594\\ \cline{3-8}
   & & Butler et al.(1997) & (A-3) & 3 &  55 Cnc b& 0.824&  14.65162 \\ \cline{6-8}
   & & & & &  Tau Boo b&3.9&3.3135\\ \cline{6-8}
   & &  & & & Ups And b & 0.69& 4.61708 \\ \cline{3-8}
   & & Johnson et al.(2008) & (A-4) & 2 &Kappa CrB b&1.8&   1191\\ \cline{6-8}
   & & & & & HD 167042  b &1.6& 416.1
\\   \hline
  (B) & Coralie &  Mayor et al. (2004) &(B-1)&16&HD 19994&2&454\\ \cline{6-8}
   & & & & &HD 65216&   1.21&   613.1\\ \cline{6-8}
   & & & & &HD 92788&3.86&  377.7\\ \cline{6-8}
   & & & & &HD 111232   &6.8&   1143 \\ \cline{6-8}
   & & & & &HD 114386   &0.99&  872\\ \cline{6-8}
   & & & & &HD 142415   &1.62&  386.3\\ \cline{6-8}
   & & & & &HD 147513   &1& 540.4\\ \cline{6-8}
   & & & & &HD 196050   &3& 1289\\ \cline{6-8}
   & & & & &HD 216437   &2.1&   1294\\ \cline{6-8}
   & & & & &HD 216770   &0.65&  118.45\\ \cline{6-8}
   & & & & &HD 6434 &0.48&  22.09\\ \cline{6-8}
   & & & & &HD 121504   &0.89&  64.6\\ \cline{6-8}
   & & & & &HD 83443 b  &0.4&   2.985625\\ \cline{6-8}
   & & & & &HD 82943 b  &1.75&  441.2\\ \cline{6-8}
   & & & & &HD 82943 c  &2.01&  219\\ \cline{6-8}
   & & & & &HD 169830 c &4.04&  2102\\  \cline{3-8}
& & Tamuz et al. (2008) &(B-2)&2&HD 4113&1.56&526.62\\\cline{6-8}
   & & & & &HD 156846   &10.45& 359.51 \\\cline{3-8}
   & & Udry  et al.(2003) &(B-3)&1&HD 73256 &1.87&  2.54858\\ \cline{3-8}
   & & Naef  et al.(2001a) & (B-4)&3& GJ 3021   &3.32&  133.82\\ \cline{6-8}
   & & & & &HD 52265    &1.13&  118.96\\ \cline{6-8}
   & & & & &HD 169830 b &2.88&  225.62 \\ \cline{3-8}
   & & Queloz  et al.(2000) &(B-5)&1 &GJ 86 &4.01&15.766 \\ \cline{3-8}
\hline
 \end{tabular}
    \end{center}
 \begin{center}
\begin{tabular}{|c|l|l|l|c|l|c|c|}\hline
 {Data Set} & Survey & Reference & Ref. ID  & N & Planet &  $M$ & $P$ \\
\hline
(B)& Coralie & Udry  et al(2000) &(B-6)& 2&HD 75289 &0.42&  3.51 \\ \cline{6-8}
   & & & & &HD 130322   &1.08&  10.724 \\ \cline{3-8}
   & &  Santos et al.(2000) & (B-7)&1 &HD 192263&0.72&24.348 \\ \cline{3-8}
   & & Santos  et al.(2001)&(B-8)& 2 &HD 28185  &5.7&   383 \\ \cline{6-8}
   & & & & &HD 213240 &4.5& 951 \\ \cline{3-8}
   & & Correia  et al.(2005) &(B-9)& 1&HD 202206 c&2.44&1383.4\\ \cline{3-8}
   & & Pepe et al. (2002) & (B-10) & 2 & HD 108147&0.4& 10.901\\\cline{6-8}
   & & & & &HD 168746   &0.23&  6.403\\ \cline{3-8}
   & & Udry  et al.(2002) & (B-11) & 4 & HD 141937&9.7&653.22 \\\cline{6-8}
   & & & & &HD 162020   &13.75& 8.428198 \\\cline{6-8}
   & & & & &HD 168443 c &18.1&  1765.8 \\\cline{6-8}
   & & & & &HD 202206 b &17.4&  255.87 \\\cline{3-8}
   & & Zucker et al. (2004) &(B-12)&1&HD 41004 A&2.3&   655\\\cline{3-8}
   & & Santos et al. (2002) &(B-13)&1&HD 41004 B&18.4&1.3283\\\cline{3-8}
   & & Eggenberger  et al.(2006) &(B-14)&1&HD 142022    &4.4&1923
   \\\hline
(C)& Elodie & Galland et al. (2005) &(C-1)&1&HD 33564&9.1&388\\\cline{3-8}
   & &  Perrier et al. (2003) &(C-2) &6&HD 8574 &2.23&  228.8\\\cline{6-8}
   & & & & &HD 23596    &7.19&  1558 \\\cline{6-8}
   & & & & &HD 33636    &9.28&  2127.7  \\\cline{6-8}
   & & & & &  HD 50554  &4.9&   1279 \\\cline{6-8}
   & & & & & HD 106252  &6.81&  1500 \\\cline{6-8}
   & & & & & HD 190228  &4.99&  1127 \\\cline{3-8}
 & & Naef et al.(2004) & (C-3) &3&HD 74156 b    &1.88&  51.65\\\cline{6-8}
  & & & & & HD 74156 c  &8.03&  2476 \\ \cline{6-8}
  & & & & & HD 145675(14 Her)   &4.64&  1773.4\\ \cline{3-8}
  & &  Naef  et al.(2003) &(C-4)&1& HD 190360 b &1.502&2891 \\ \cline{3-8}
  & & Moutou  et al.(2006) &(C-5)&1&HD 185269   &0.94&  6.838 \\ \cline{3-8}
  & & da. Silva  et al.(2006) &(C-6)&1& HD 118203&2.13&6.1335 \\ \cline{3-8}
 & &Mayor \& Queloz (1995) & (C-7)  & 1 & 51 Peg & 0.468 & 4.23077  \\
 \hline
(D) &HARPS & Udry et al. (2006) &(D-1)&1& HD 4308&0.047&15.56\\ \hline

\end{tabular}
    \end{center}

 \begin{center}
\begin{tabular}{|c|l|l|l|c|l|c|c|}\hline
 {Data Set} & Survey & Reference & Ref. ID  & N & Planet &  $M$ & $P$ \\
\hline
(D) &HARPS & Moutou et al.(2005) &(D-2)&3&HD 2638
&0.48&  3.4442\\ \cline{6-8}
  & & & & & HD 27894    &0.62&  17.991 \\ \cline{6-8}
  & & & & & HD 63454    &0.38&  2.81782 \\ \cline{3-8}
  & & Pepe et al. (2004) & (D-3) &1&HD 330075 b &0.76&  3.369\\  \cline{3-8}
  & & Lo Curto et al.(2006) & (D-4)&1& HD 212301 &0.45& 2.457 \\  \cline{3-8}
  & & Lovis  et al.(2005) & (D-5)&3& HD 93083   &0.37&  143.58 \\  \cline{6-8}
  & & & & & HD 10193    &0.3&   70.46 \\  \cline{6-8}
  & & & & & HD 102117 &0.172&   20.67 \\  \cline{3-8}
  & &  Santos  et al.(2004) & (D-6)&1&HD 160691 d&0.044&9.55 \\\cline{3-8}
  & & Pepe  et al.(2007) &(D-7)&1&HD 160691 e   &0.5219& 310.55\\\cline{3-8}
  & & Bonfils  et al.(2005) & (D-8) &1& Gl 581 b&0.0492& 5.3683\\\cline{3-8}
  & & Udry  et al.(2007) & (D-9) & 2 & Gl 581 c &0.0158&12.932\\\cline{6-8}
  & & & & &Gl 581 d &0.0243&    83.6 \\\cline{3-8}
  & & Bonfils et al.(2007) &(D-10)&1& GJ 674    &0.037& 4.6938 \\\cline{3-8}
  & & Melo et al.(2007) &(D-11)&1& HD 219828    &0.066& 3.8335 \\\cline{3-8}
  & & Santos et al. (2007)&(D-12)&1&HD 171028   &1.83&  538\\\cline{3-8}
  & & Naef et al. (2007) &(D-13)&3&HD 100777&1.16&383.7\\\cline{6-8}
  & & & & & HD 190647   &1.9&   1038.1 \\ \cline{6-8}
  & & & & & HD 221287   &3.09&  456.1\\ \cline{3-8}
  & & Lovis et al. (2006)&(D-14)&3& HD 69830 b  &0.033& 8.667\\\cline{6-8}
  & & & & & HD 69830 c  &0.038& 31.56 \\\cline{6-8}
  & & & & & HD 69830 d  &0.058& 197 \\  \hline
(E)& N2K & Wright et al. (2007) &(E-1)&3&HIP 14810 b&3.84&6.6742\\\cline{6-8}
 & & & & & HIP 14810 c  &0.76&  95.2914 \\\cline{6-8}
 & & & & & HD 154345 b  &2.03&  10900 \\\cline{3-8}
& & Johnson et al. (2006)& (E-2)&3&HD 33283 &0.33   &18.179\\\cline{6-8}
& & & & &HD 86081   &1.5    &2.1375 \\\cline{6-8}
& & & & &HD 224693  &0.71   &26.73 \\\cline{3-8}
& & Fischer  et al.(2006) &(E-3) &2&HD 149143   &1.33&  4.072\\ \cline{6-8}
& & & & & HD 109749 &0.28&  5.24 \\ \cline{3-8}
& &Fischer  et al.(2007) &(E-4)&5&HD 11506  &4.85&  1280 \\ \cline{6-8}
& & & & &HD 125612  &3.2&   502 \\ \cline{6-8}
& & & & &HD 231701  &1.78&  141.6 \\ \hline
\end{tabular}
    \end{center}

\begin{center}
\begin{tabular}{|c|l|l|l|c|l|c|c|}\hline
 {Data Set} & Survey & Reference & Ref. ID  & N & Planet &  $M$ & $P$ \\
\hline
(E)& N2K  & && &HD 170469   &0.67&  1145 \\ \cline{6-8}
& & & & &HD 17156 b &3.111& 21.21725\\ \cline{3-8}
& & Sato et al.(2005) & (E-5) &1&HD 149026 &0.36&  2.8766\\ \cline{3-8}
& &Fischer et al.(2005) & (E-6)  & 1 & HD 88133 & 0.22 & 3.41
\\ \hline
(F) & Keck & Butler et al. (2003) & (F-1)&6&HD 108874 b&1.36&395.4\\ \cline{6-8}
& & & & & HD 114729 &0.82&  1131.478\\  \cline{6-8}
& & & & & HD 72659  &2.96&  3177.4\\  \cline{6-8}
& & & & & HD 128311 b   &2.18&  448.6 \\  \cline{6-8}
& & & & & HD 49674  &0.115& 4.9437\\  \cline{6-8}
& & & & & HD 37124 c    &0.683& 2295\\  \cline{3-8}
& &Marcy et al.(2005) &(F-2)&5&HD 183263    &3.69&  634.23 \\  \cline{6-8}
& & & & &HD 117207  &2.06&  2627.08 \\  \cline{6-8}
& & & & &HD 188015  &1.26&  456.46\\  \cline{6-8}
& & & & &HD 45350   &1.79&  890.76 \\  \cline{6-8}
& & & & &HD 99492   &0.109& 17.0431 \\  \cline{3-8}
& & Robinson  et al.(2007)& (F-3)&2&HD 75898 b  &1.48&  204.2 \\  \cline{6-8}
& & & & &HD 5319 b      &1.94&  675\\  \cline{3-8}
& &Butler  et al.(1998)&(F-4)&1&HD 187123 b &0.52&  3.097 \\  \cline{3-8}
& & Marcy  et al.(1999)&(F-5)&2&HD 210277   &1.23&  442.1\\\cline{6-8}
& & & & &HD 168443 b    &8.02&  58.11289\\ \cline{3-8}
& & Johnson et al.(2007) &(F-6)&1&GJ 317    &1.2&   692.9\\\cline{3-8}
& &Vogt  et al.(2000) & (F-7) &6&HD 10697&6.12&1077.906 \\\cline{6-8}
& & & & &HD 37124 b &0.61&  154.46 \\\cline{6-8}
& & & & &HD 134987  &1.58&  260\\\cline{6-8}
& & & & &HD 177830  &1.28&  391\\\cline{6-8}
& & & & &HD 192263  &0.72&  24.348 \\\cline{6-8}
& & & & &HD 22258      &5.11&   572 \\\cline{3-8}
 & &
Butler  et al.(2006b)&(F-8)& 1   &GJ 849    &0.82&  1890\\  \cline{3-8}
&&  Vogt et al. (2002) &(F-9) & 5 &HD 4203  &1.65&  400.944\\  \cline{6-8}
& & &&&HD 4208  &0.8&   812.197\\\cline{6-8}
& & &&&HD 33636 &9.28&  2127.7 \\\cline{6-8}
& & &&&HD 68988 &1.9&   6.276\\\cline{6-8}
& & &&& HD 114783   &0.99&  501\\
\hline
\end{tabular}
    \end{center}

\begin{center}
\begin{tabular}{|c|l|l|l|c|l|c|c|}\hline
 {Data Set} & Survey & Reference & Ref. ID  & N & Planet &  $M$ & $P$ \\
\hline
(F)& Keck &
Marcy  et al.(2001a) &(F-10)&1&HD 168443 c  &18.1&  1765.8 \\\cline{3-8}
&&Rivera et al. (2005) &(F-11) &1&GJ 876 d  &0.018& 1.93776 \\\cline{3-8}
&& Vogt  et al.(2005) & (F-12) &7& HD 128311 c &3.21& 919\\ \cline{6-8}
& & &&&HD 50449     &1.71&  2582.7\\\cline{6-8}
& & &&&HD 37124 d   &0.6&   843.6\\\cline{6-8}
& & &&&HD 190360 c  &0.057& 17.1\\\cline{6-8}
& & &&& HD 108874c      &1.018& 1605.8\\\cline{6-8}
& & &&& HD 37124 c  &0.683& 2295\\\cline{6-8}
& & &&& HD 217107 c     &2.5&   3352 \\\cline{3-8}
&& Butler  et al.(2004)&(F-13)&1& GJ 436&0.072&2.64385 \\\cline{3-8}
&& Butler  et al.(2006a)&(F-14)& 5&  HD 11964 b &0.11&  37.82\\\cline{6-8}
& & &&&HD 66428 &2.82&  1973\\\cline{6-8}
& & &&&HD 99109 &0.502& 439.3\\\cline{6-8}
& & &&&HD 107148    &0.21&  48.056\\\cline{6-8}
& & &&&HD 164922    &0.36&  1155\\\cline{3-8}
& & Marcy  et al.(2000) &(F-15) &2 &HD 16141    &0.23&  75.56\\\cline{6-8}
& & &&&HD 46375 &0.249& 3.024\\\cline{3-8}
& & Butler  et al.(2000) &(F-16) &2&HD 52265    &1.13&  118.96\\\cline{6-8}
& & &&&BD 103166    &0.48&  3.488\\ \hline
(G)& AAPS& Tinney  et al.(2003)&(G-1) &4& HD 73526 b    &2.9&   188.3\\\cline{6-8}
& & &&&HD 76700 &0.197& 3.971\\\cline{6-8}
& & &&&HD 30177 &9.17&  2819.654\\\cline{6-8}
& & &&&HD 2039 &4.85&   1192.582\\\cline{3-8}
& &Jones  et al.(2002b)&(G-2) &3&  HD 196050    &3& 1289 \\\cline{6-8}
& & &&&HD 216437    &2.1&   1294\\\cline{6-8}
& & &&&HD 160691 c  &3.1&   2986\\\cline{3-8}
&& Jones  et al.(2002a)& (G-3) &1&HD 39091  &10.35& 2063.818\\\cline{3-8}
&& Tinney  et al.(2002)&(G-4)& 2 &HD 142    &1& 337.112\\\cline{6-8}
&& & &&HD 23079 &2.61&  738.459 \\\cline{3-8}
&& Jones  et al.(2006) & (G-5) &2&HD 187085 &0.75&  986\\\cline{6-8}
& & &&&HD 20782 &1.8&   585.86\\\hline
\end{tabular}
    \end{center}
\begin{center}
\begin{tabular}{|c|l|l|l|c|l|c|c|}\hline
 {Data Set} & Survey & Reference & Ref. ID  & N & Planet &  $M$ & $P$ \\
\hline
(G)& AAPS&
O'Toole et al.(2007)& (G-6) &2&HD 23127     &1.5&   1214\\\cline{6-8}
& & &&&HD 159868    &1.7&   986\\\cline{3-8}
&& Butler  et al.(2001)& (G-7) &2&HD 160691 b   &1.67&  654.5\\\cline{6-8}
& & &&&HD 27442     &1.28&  423.841\\\cline{3-8}
&& Carter  et al.(2003)& (G-8)&1&HD 70642   &2& 2231\\\cline{3-8}
&& Tinney  et al.(2001)&(G-9)&1&HD 179949   &0.95&  3.0925\\\cline{3-8}
&&Jones  et al.(2003)&(G-10)&1&Tau Gruis b  &1.49&  1442.919\\\cline{3-8}
&&Tinney  et al.(2006)&(G-11)&1&HD73526 c   &1.6&   416.1\\\cline{3-8}
&& Tinney et al.(2005) & (G-12) & 3 & HD 117618 & 0.19 &    52.2\\ \cline{6-8}
&  &  &  &  & HD 208487 & 0.45 & 123 \\ \cline{6-8}
   & & & & & HD 102117 & 0.172 & 20.67 \\ \hline
(H)& Others&  Fischer et al. (2001)&(H-1)& 3&HD 12661 b&2.3&    263.6\\\cline{6-8}
& & && &HD 92788    &3.86&  377.7\\\cline{6-8}
& & && &HD 38529 b  &0.78&  14.30\\\cline{3-8}
& &Marcy  et al.(2001b)&(H-2)&1& GJ 876 c&0.56&30.1  \\\cline{3-8}
& & Fischer et al. (2003)&(H-3) &3&HD 40979 &3.32&  267.2\\\cline{6-8}
& & &&&HD 12661 c   &1.57&  1444.5\\\cline{6-8}
& & &&&HD 38529 c   &12.7&  2174.3\\\cline{3-8}
&& Delfosse  et al.(1998)&(H-4)& 1&GJ 876 b &1.935& 60.94\\\cline{3-8}
&& Cochran et al. (1997) &(H-5)& 1&16 Cyg B b&1.68&799.5\\\cline{3-8}
&& Sozzetti et al. (2006)& (H-6)& 1& HD 81040   &6.86&  1001.7\\\cline{3-8}
&& Naef  et al.(2001b)& (H-7)& 1&HD 80606 b &3.41&  111.78\\\cline{3-8}
&& Fischer et al. (1999)& (H-8)&2&HD 195019 A b&3.7&    18.20163\\\cline{6-8}
& & &&&HD 217107    &1.33&  7.12689\\\cline{3-8}
&& Butler  et al.(1999) &(H-9) &2&Ups And c &1.98&  241.52 \\\cline{6-8}
& & &&&Ups And d et al. &3.95&  1274.6\\\cline{3-8}
& & Fischer  et al.(2002b) &(H-10)& 1&HD 136118 &11.9&  1209\\\cline{3-8}
&& Korzennik  et al.(2000) &(H-11) &1&HD 89744&7.99&    256.605\\\hline
\end{tabular}
    \end{center}

\clearpage
\section*{Appendix B}

The derivation of the log likelihood $\ln L$ of coupled mass-period
functions is shown here.
As in Tabachnik and Tremaine (2002), the likelihood function is
\beqn
L &=& \prod_{i=1}^N  n(x_i,y_i)
\exp\Big[-\int_D n(x,y)dxdy\Big], \non \\
&=&\prod_{i=1}^N c  N_{s} f_{XY}(x_i,y_i)
\exp\left[-cN_s \int_{y_{\rm min}}
^{\tilde{u}}
\int_{v+y/3}^{x_{\rm max}} f_{XY}(x,y)dxdy\right].\label{eq:exp1}
\eeqn


By the other equations in \S 6, we have
\beqn
& &\int_{y_{\rm min}}^{\tilde{u}}\int_{v+y/3}^{x_{\rm max}}
 f_{XY}(x,y)dxdy
 =\int_{y_{\rm min}}^{\tilde{u}}
\int_{v+y/3}^{x_{\rm max}}\frac{\partial^2 C (F_X(x),F_Y(y);\theta)}
      {\partial F_X \partial F_Y} f_X(x) f_Y(y) dxdy, \non \\
&=& \int_{y_{\rm min}}^{\tilde{u}}
\left\{\frac{\partial C (F_X(x_{\rm max}),F_Y(y);\theta)}{\partial F_Y}
-\frac{\partial C (F_X(v+y/3),F_Y(y);\theta)}{\partial F_Y} \right\}
f_Y(y) dy, \non \\
&=& C (F_X(x_{\rm max}),F_Y(\tilde{u});\theta)- C(F_X(x_{\rm max}),
F_Y(y_{\rm min});\theta)-\int_{y_{\rm min}}^{\tilde{u}}
\frac{\partial C (F_X(v+y/3),F_Y(y);\theta)}{\partial F_Y}
f_Y(y) dy, \non \\
&=& F_Y(\tilde{u})-\int_{y_{\rm min}}^{\tilde{u}}
\frac{[e^{-\theta (F_Y(y)+F_X(v+y/3))}-
 e^{-\theta F_Y(y)}]}
 {e^{-\theta}+e^{-\theta (F_Y(y)+F_X(v+y/3))}-e^{-\theta F_Y(y)}-
 e^{-\theta F_X(v+y/3)}}f_Y(y) dy,\label{eq:exp2}
\eeqn
\
 where  $C(F_X(x_{\rm max}),F_Y(y_{\rm min});\theta)=C(1,0,\theta)=0 $ ,and
$C (F_X(x_{\rm max}),F_Y(\tilde{u});\theta)=C(1,F_Y(\tilde{u});\theta)
=F_Y(\tilde{u}).$
 From Eq.(\ref{eq:exp1}) and Eq.(\ref{eq:exp2}), we have
\beqn
 & & \exp\Big[-\int_D n(x,y)dxdy\Big] \non \\
 &= & \exp\left\{-c N_{s}\Big(F_Y(\tilde{u})-
 \int_{y_{\rm min}}^{\tilde{u}} \frac{[e^{-\theta (F_Y(y)+F_X(v+y/3))}-
 e^{-\theta F_Y(y)}]f_Y(y)}
 {e^{-\theta}+e^{-\theta (F_Y(y)+F_X(v+y/3))}-e^{-\theta F_Y(y)}-
 e^{-\theta F_X(v+y/3)}}dy\Big)\right\},
\end{eqnarray}
and the log-likelihood is
 \beqn
\ln L &=& N\ln(c N_{s}) -\alpha \sum_{i=1}^N x_i-\beta \sum_{i=1}^N y_i+
 N\ln \Big(\frac{\alpha}{e^{-\alpha x_{\rm min}}-e^{-\alpha x_{\rm max}}}\Big)
 +N\ln \Big(\frac{\beta}{e^{-\beta y_{\rm min}}-e^{-\beta y_{\rm max}}}\Big)
\non \\
& &+\sum_{i=1}^N \ln\Big(C_{u_1u_2}(F_X(x_i),F_Y(y_i);\theta)\Big) - c N_{s}
I(\alpha,\beta,u,v),\label{eq:ln_l2}
  \eeqn
where
\beqn
 C_{u_1u_2}(F_X(x_i),F_Y(y_i);\theta) &=& \frac{-\theta(e^{-\theta}-1)
 e^{-\theta F_X(x_i)}e^{-\theta F_Y(y_i)}}
 {\Big[e^{-\theta}-1+(e^{-\theta F_X(x_i)}-1)(e^{-\theta F_Y(y_i)}-1)\Big]^2};
 \eeqn
 \beqn
I(\alpha,\beta,u,v)
 &=&  F_Y(\tilde{u})-\int_{y_{\rm min}}^{\tilde{u}}
 \frac{[e^{-\theta (F_Y(y)+F_X(v+y/3))}-e^{-\theta F_Y(y)}]f_Y(y)}
 {e^{-\theta}+e^{-\theta (F_Y(y)+F_X(v+y/3))}-e^{-\theta F_Y(y)}
 -e^{-\theta F_X(v+y/3))}}dy.
 \eeqn

By ${\pa \ln L}/{\pa c}=0$,
we obtain
$c=N/[N_{s} I(\alpha,\beta,u,v)]$.
It is then substituted
into Eq.(\ref{eq:ln_l2})
and we have
\beqn
 \ln L &=& N\ln(N)-N\ln (I) -\alpha \sum_{i=1}^N x_i-\beta \sum_{i=1}^N y_i+
 N\ln \Big(\frac{\alpha}{e^{-\alpha x_{\rm min}}-e^{-\alpha x_{\rm max}}}\Big)
\non \\
& &+N\ln \Big(\frac{\beta}{e^{-\beta y_{\rm min}}-e^{-\beta y_{\rm max}}}\Big)
+\sum_{i=1}^N \ln\Big(C_{u_1u_2}(F_X(x_i),F_Y(y_i);\theta)\Big)-N.
\label{eq:ln_l3}
  \eeqn


\clearpage

\begin{figure}
\epsscale{1.0}
\plotone{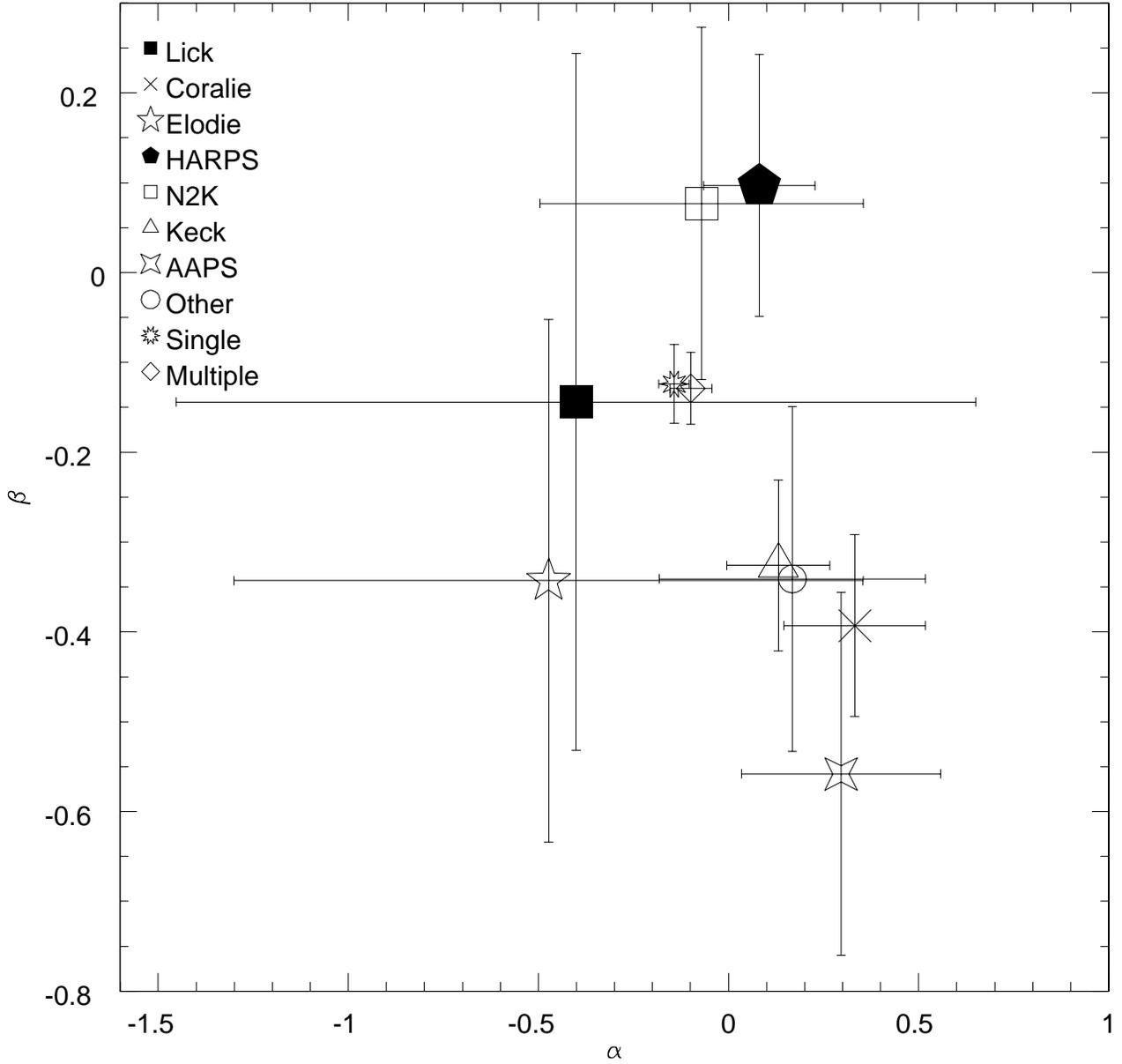}
\caption{The estimators of the exponents, $\alpha$ and $\beta$, with
error bars for all surveys listed in Table 1,
and also the results of Multiple Surveys.}
\end{figure}

\clearpage
\begin{figure}
\epsscale{1.0}
\plotone{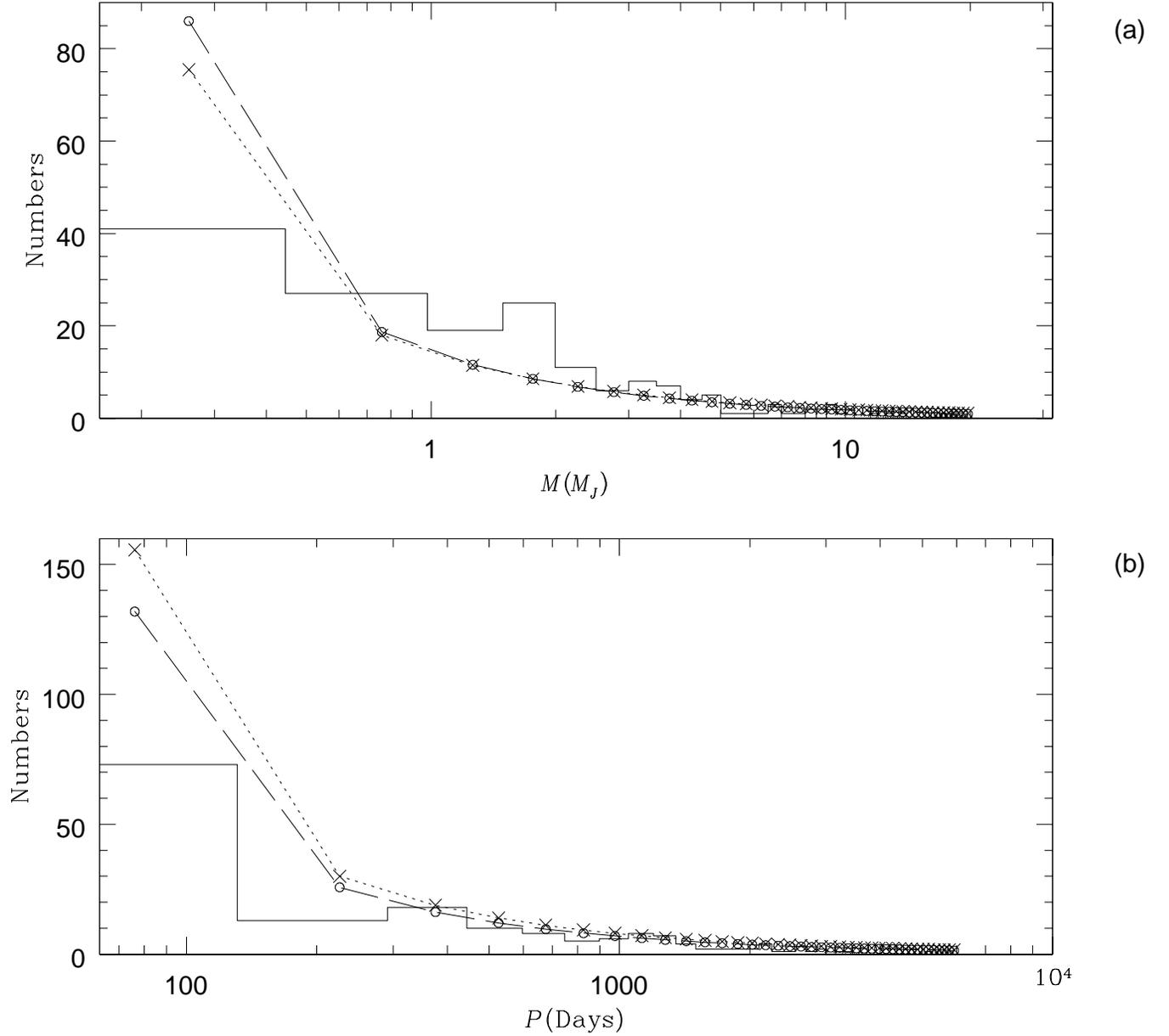}
\caption{The distributions of exoplanets in $M$ and $P$ space.
The solid lines are the histograms of 175 exoplanets.
By Eq.(24),
the crosses with dotted curves are the results of Single Imaginary
Survey, and the circles with dashed curves are the results of Multiple
Surveys.
}
\end{figure}


\clearpage
\begin{figure}
\epsscale{1.0}
\plotone{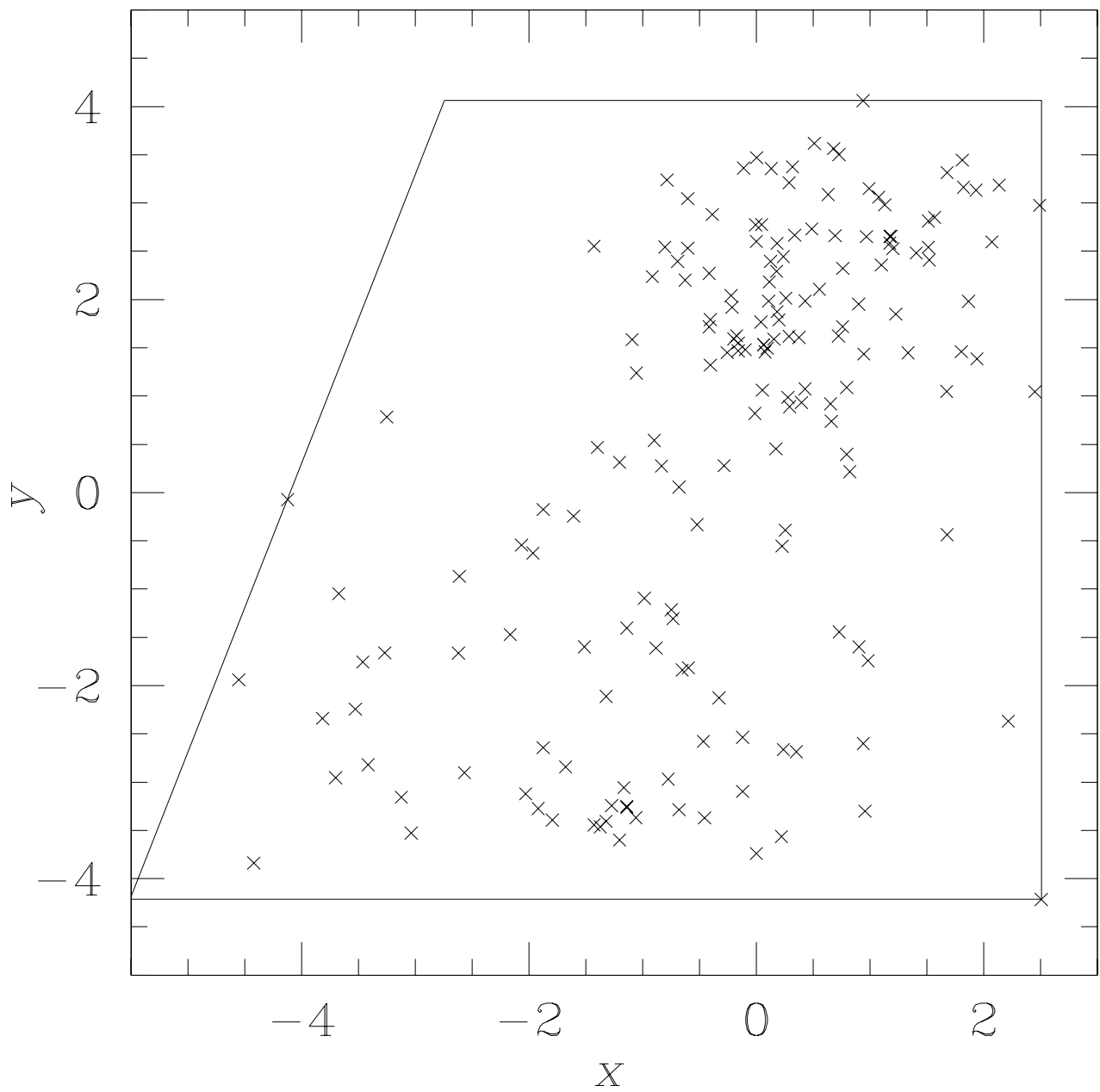}
\caption{The distribution of 175 exoplanets in $x-y$ space.
The solid lines indicate Domain D.}
\end{figure}

\clearpage
\begin{figure}
\epsscale{1.0}
\plotone{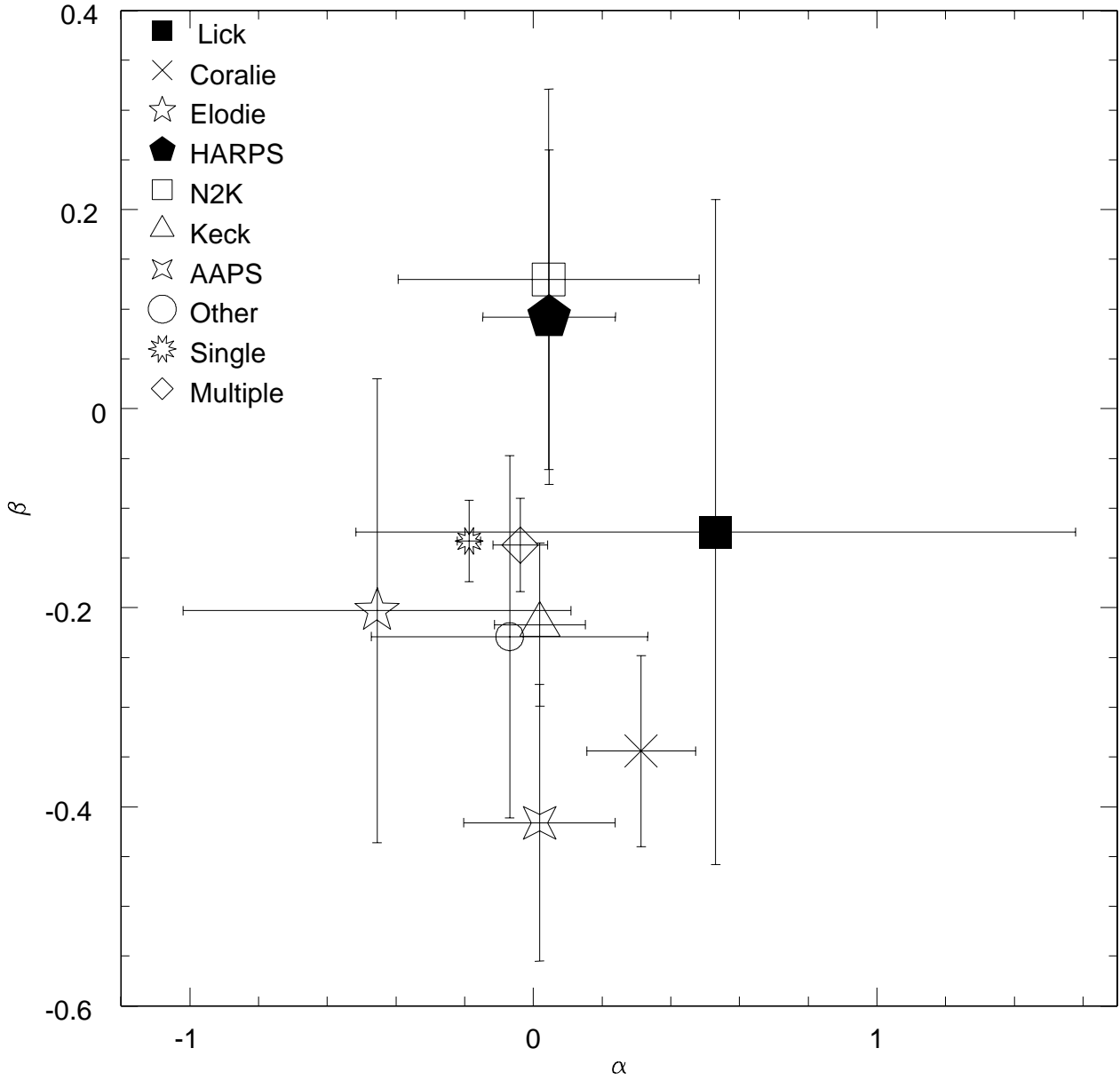}
\caption{The
estimators of the exponents, $\alpha$ and $\beta$, with
error bars for all surveys listed in Table 4.
}
\end{figure}

\clearpage
\begin{figure}
\epsscale{1.0}
\plotone{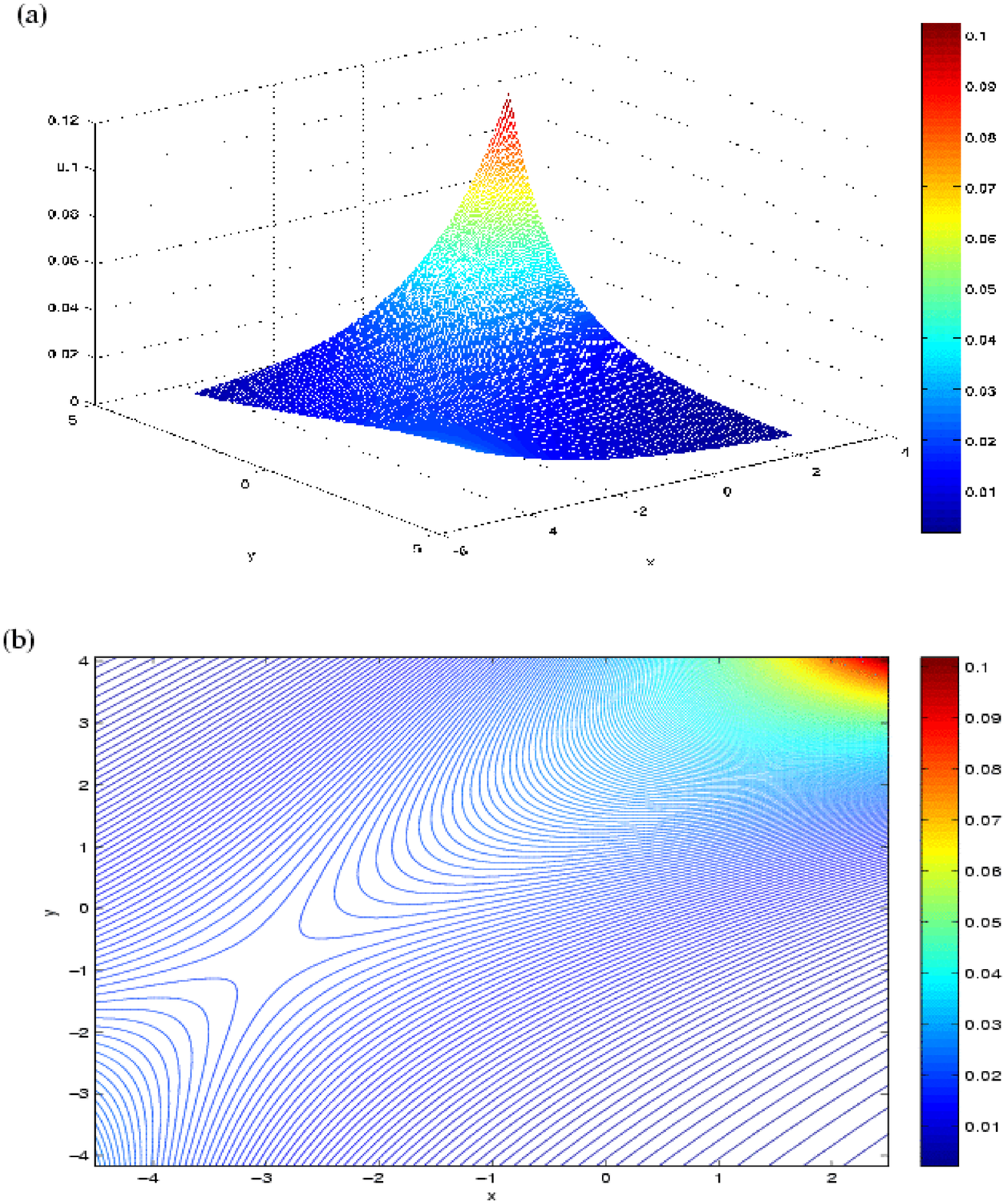}
\caption{(a) The coupled probability density function in $x-y$ space,
i.e. $f_{XY}(x,y)$, from the results of Multiple Surveys.
(b) The corresponding contour plot.
}
\end{figure}

\clearpage
\begin{figure}
\epsscale{1.0}
\plotone{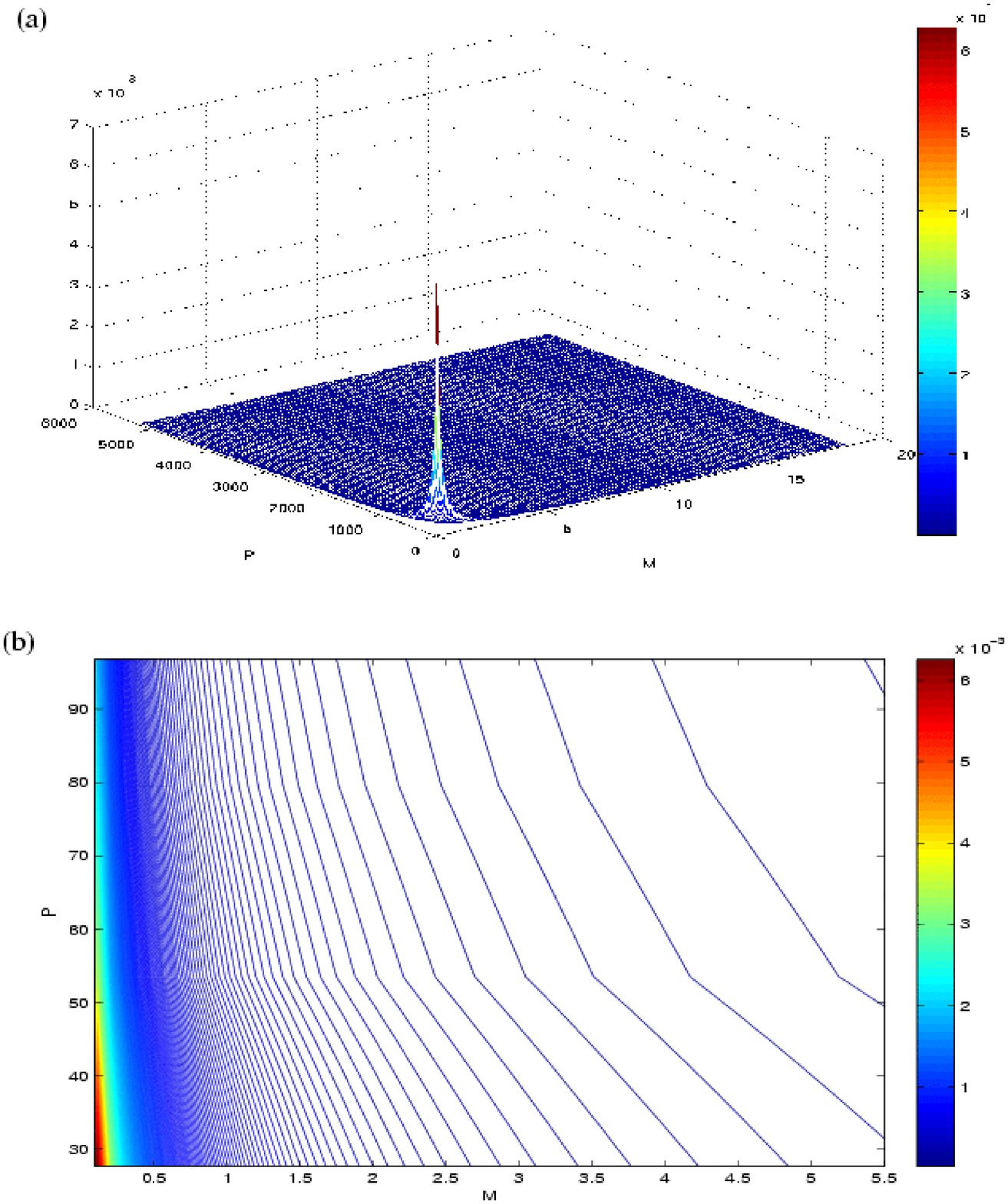}
\caption{(a) The coupled probability density function in $M-P$ space,
i.e. $f_{MP}(M,P)$, from the results of Multiple Surveys.
(b) The corresponding contour plot.
}
\end{figure}


\clearpage
\begin{figure}
\epsscale{1.0}
\plotone{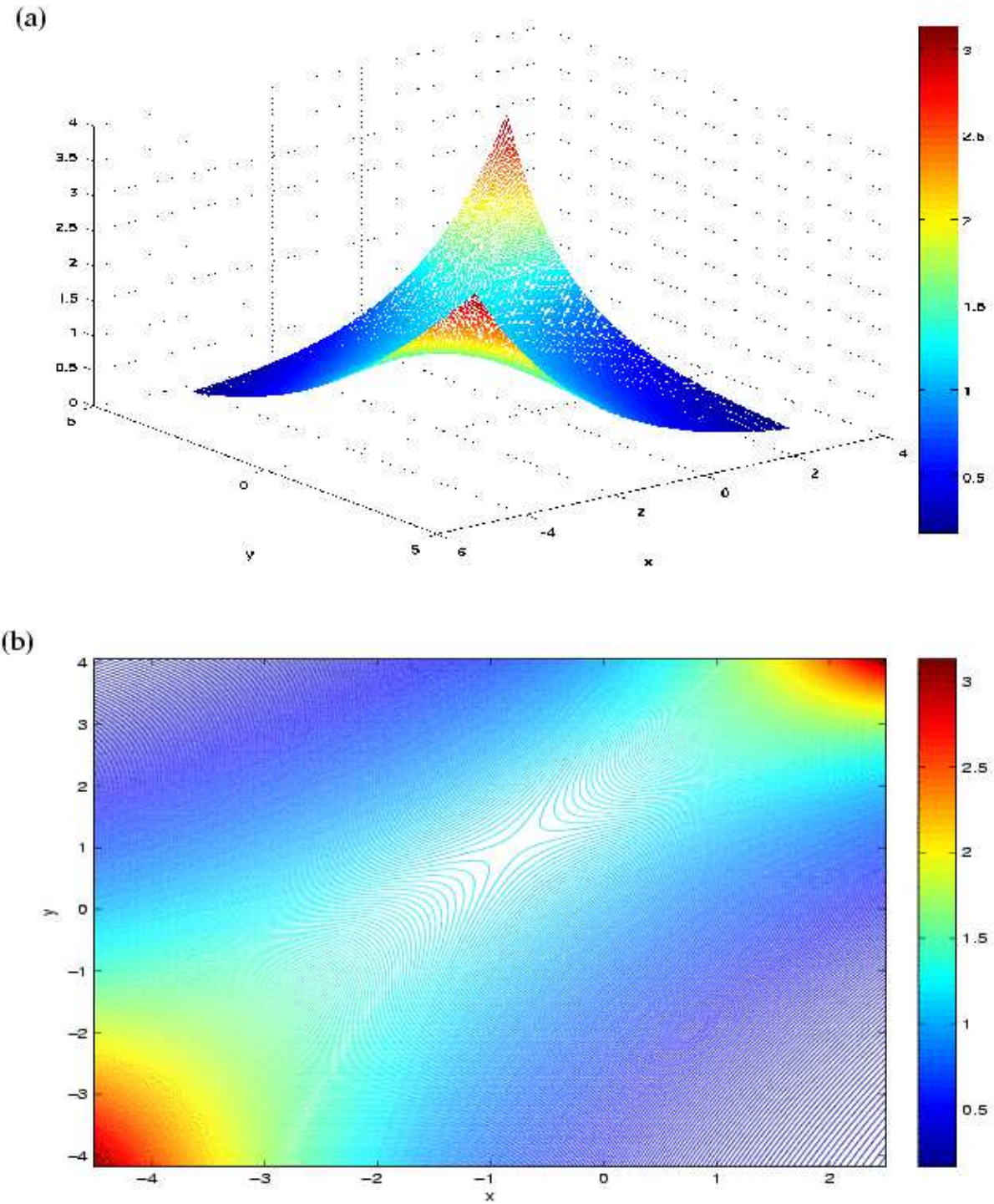}
\caption{(a) The three-dimensional plot of the coupling factor
for the results of Multiple Surveys.
(b) The corresponding contour plot.}
\end{figure}


\clearpage
\begin{figure}
\epsscale{1.0}
\plotone{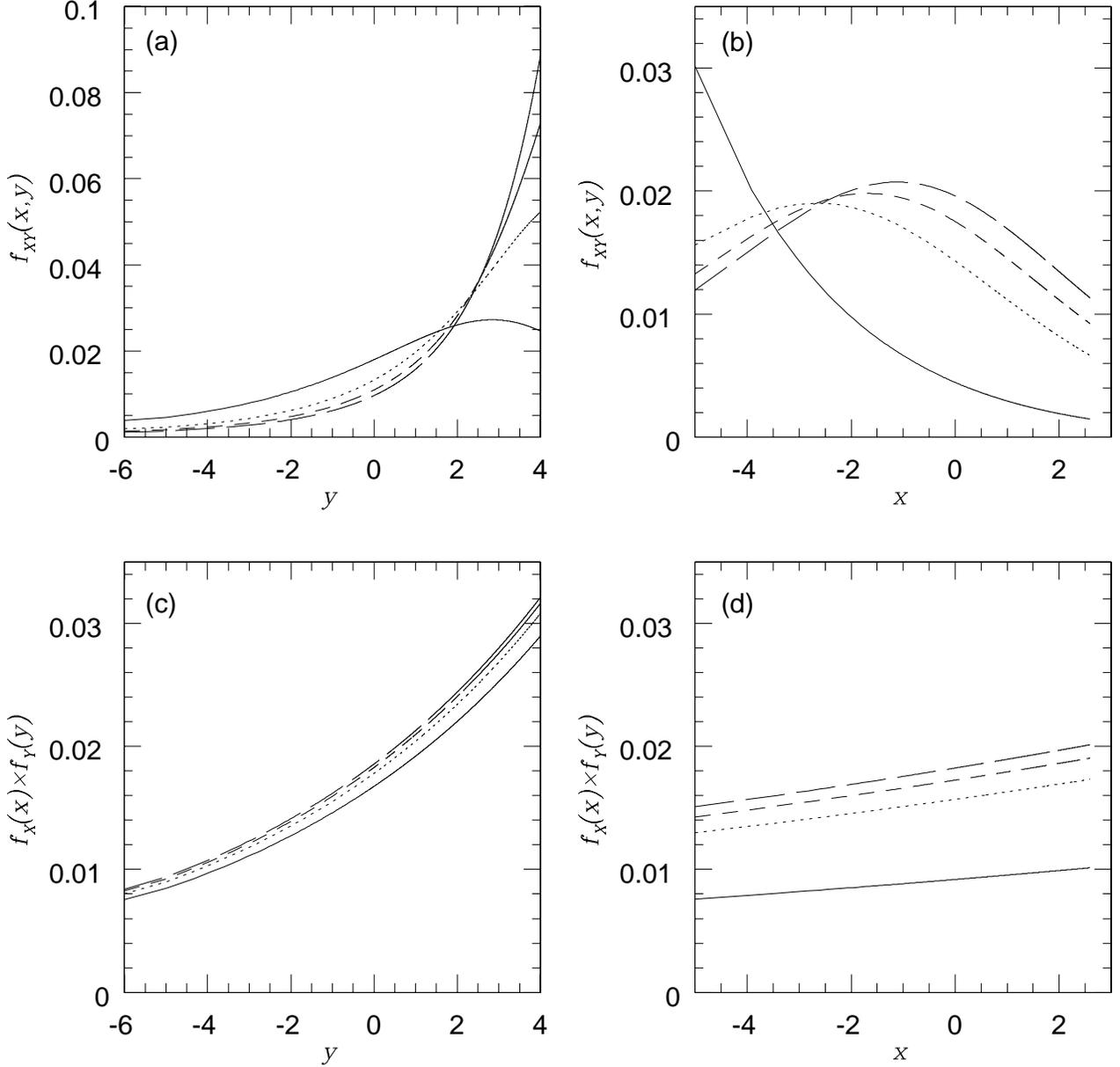}
\caption{The mass and period functions in $x-y$ space
for the results of Multiple Surveys.
(a) The period functions of $x$=ln($1 M_J/M_0$) (solid curve),
$x$=ln($5 M_J/M_0$) (dotted curve), $x$=ln($10 M_J/M_0$) (short
dashed curve), and  $x$=ln($15 M_J/M_0$)  (long dashed curve).
(b) The mass functions of $y$=ln($1 {\rm day}/P_0$) (solid curve),
$y$=ln($50 {\rm days}/P_0$) (dotted curve),
$y$=ln($100 {\rm days}/P_0$) (short dashed curve), and
$y$=ln($150 {\rm days}/P_0$) (long dashed curve).
(c) The independent period functions of $x$=ln($1 M_J/M_0$) (solid curve),
$x$=ln($5 M_J/M_0$) (dotted curve), $x$=ln($10 M_J/M_0$) (short
dashed curve), and  $x$=ln($15 M_J/M_0$)  (long dashed curve).
(d) The independent mass functions of $y$=ln($1 {\rm day}/P_0$) (solid curve),
$y$=ln($50 {\rm days}/P_0$) (dotted curve),
$y$=ln($100 {\rm days}/P_0$) (short dashed curve), and
$y$=ln($150 {\rm days}/P_0$) (long dashed curve).
}
\end{figure}

\clearpage
\begin{figure}
\epsscale{1.0}
\plotone{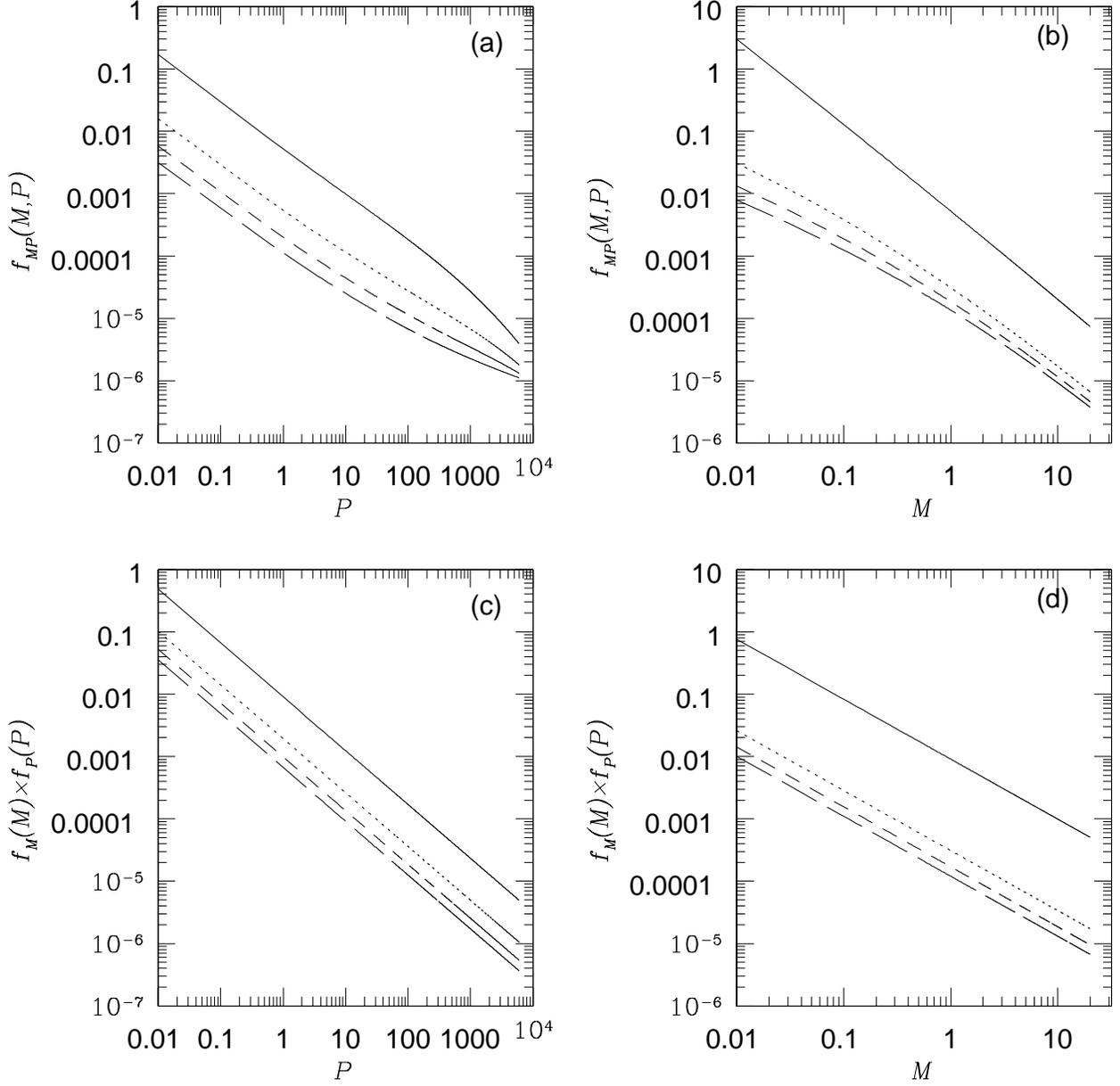}
\caption{The mass and period functions in $M-P$ space
for the results of Multiple Surveys.
(a) The period functions of $M=1
M_J$ (solid curve), $M=5 M_J$ (dotted curve), $M=10 M_J$ (short
dashed curve), and $M=15 M_J$ (long dashed curve). (b) The mass
functions of $P=1$ day (solid curve), $P=50$ days (dotted curve),
$P=100$ days (short dashed curve), and $P=150$ days (long dashed
curve). (c) The independent period functions of $M=1 M_J$ (solid
curve), $M=5 M_J$ (dotted curve), $M=10 M_J$ (short dashed curve),
and $M=15 M_J$ (long dashed curve). (d) The independent mass
functions of $P=1$ day (solid curve), $P=50$ days (dotted curve),
$P=100$ days (short dashed curve), and $P=150$ days (long dashed
curve).
}
\end{figure}

\end{document}